\documentclass[a4paper,11pt,article]{article}
\usepackage{textcomp}
\usepackage[english]{babel}
\usepackage{graphicx}

\usepackage{booktabs}
\usepackage{graphics}
\usepackage{subcaption}
\usepackage{mwe}
\usepackage[utopia]{mathdesign}
\usepackage{bbm}
\usepackage[table]{xcolor}
\usepackage{booktabs}
\usepackage{dcolumn}
\usepackage{dsfont}
\usepackage{setspace}
\usepackage{comment}
\usepackage{tikz}
\usetikzlibrary{positioning,arrows.meta}
\usetikzlibrary{trees}
\usepackage{pgfplots}
\pgfplotsset{width=10cm,compat=1.9,title style={at={(.35,-.28)}}}
\usetikzlibrary{calc}
\usepackage{float} %package to force figures to stay in place
\usepackage{mathpazo} % math & rm
\usepackage[T1]{fontenc}
\normalfont

\usepackage{amsthm}
\usepackage{amsmath}
\usepackage{gensymb}
\usepackage{wasysym}
\usepackage{marvosym}
\usepackage[margin=1 in]{geometry}
\usepackage{tikz}
\usetikzlibrary{positioning}
\usepackage[titletoc,title]{appendix}
\usepackage{longtable}
\usepackage{multirow}
\usepackage{lscape}
\usepackage{bigstrut}
\usepackage[sort&compress]{natbib}
\usepackage{longtable}

\theoremstyle{plain}

\theoremstyle{definition}
\newtheorem{fnd}{Finding}
\theoremstyle{definition}
\newtheorem{hyp}{Hypothesis}
\newtheorem*{hyp*}{Hypothesis}

\theoremstyle{definition}

\usepackage{hyperref}
\hypersetup{
    colorlinks=true,
    linkcolor=blue,
    citecolor=blue,
    urlcolor=cyan,
    pdftitle={Sharelatex Example},
    bookmarks=true,
    breaklinks=true
}

\usepackage{enumitem}

\begin{document}
\newgeometry{top=0.1cm} % Reduce top margin just for the title page
\title{Position Uncertainty in a Prisoner's Dilemma Game: An Experiment}
\author{Chowdhury Mohammad Sakib Anwar\footnote{Business School, Faculty of Business, Law \& Digital Technology,
University of Winchester, SO22 4NR, Winchester, U.K., {\Letter}: sakib.anwar@winchester.ac.uk, ORCID: 0000-0002-0223-7963}
\and
Konstantinos Georgalos\footnote{Department of Economics, Lancaster University Management School, LA1 4YX, Lancaster, U.K.,{\Letter}: k.georgalos@lancaster.ac.uk, ORCID: 0000-0002-2655-4226
}.
%The instructions, data, codes and replication files can be found at \url{https://osf.io/br5uy}
\and 
Sonali SenGupta \footnote{Department of Economics, Queen's Business School, BT9 5EE, Belfast, U.K., {\Letter}: s.sengupta@qub.ac.uk, ORCID: 0000-0002-8731-463X\newline
The study has been pre-registered at the Open Science Framework (OSF) registries (\url{https://osf.io/fzm5t}).\newline We are grateful to the Department of Economics at the Lancaster University Management School for providing the funds for the experimental sessions.}}

\maketitle
\vspace{-3em}
\begin{abstract}
\setstretch{1.4}
\citet{gallice2019co} present a natural environment that sustains full cooperation in one-shot social dilemmas among a finite number of self-interested agents. They demonstrate that in a sequential public goods game, where agents lack knowledge of their position in the sequence but can observe some predecessors’ actions, full contribution emerges in equilibrium due to agents’ incentive to induce potential successors to follow suit. Furthermore, they show that this principle extends to a number of social dilemmas, with the prominent example that of the prisoner's dilemma. In this study, we experimentally test the theoretical predictions of this model in a multi-player prisoner's dilemma environment, where subjects are not aware of their position in the sequence and receive only partial information on past cooperating actions. We test the predictions of the model, and through rigorous structural econometric analysis, we test the descriptive capacity of the model against alternative behavioural strategies, such as conditional cooperation, altruistic play and free-riding behaviour. We find that the majority resorts to free-riding behaviour, around 30\% is classified as \citet{gallice2019co} types, followed by those with social preference considerations and the unconditional altruists.  


\end{abstract}
\pagebreak
\restoregeometry
\section{Introduction}

An interesting question in the literature of environmental economics, which still remains (somewhat) unanswered, is how the nations, acting non-cooperatively, can deal with and achieve the goal of global emission abatement. Similar to public good provisions, in public bad situations such as abatement, the non-cooperative interaction among participants typically results in low levels of provision or abatement. The seminal paper by \cite{Barrett1994} provides a simple abatement game (which acts as a prisoner's dilemma) where each country must decide whether to adopt eco-friendly policies \textit{(cooperate)} or prioritize short-term economic interests \textit{(defect)}. Each country observes the choices of those preceding them \textit{(sequential decision-making)}, and these interactions capture the global environmental impact and economic consequences of climate-related policies. If all countries cooperate, there is a better chance of achieving global climate goals and minimizing environmental damage.
However, if one or more countries defect by not taking sufficient action, it may undermine the collective effort and exacerbate global environmental problems. Similar difficulties also exist in situations involving management of shared environmental resources such as forests, fishery or water basin, where agents learn by observing the actions of others preceding them (observational learning), and may strategically anticipate the actions of future agents (\citealt{Banerjee1992}; \citealt{CelenKariv2004}). This captures the essence of social dilemmas, where the optimal outcome requires coordinated efforts from all participants.
It also explores how decision-making sequence and observations of others' behaviour influence individual choices in social dilemma contexts \citep{suleiman1994position}. 

Cooperation in social dilemma situations is feasible, in the case of a repeated strategic interaction over an infinite horizon (\citealt{Friedman1971}; \citealt{DuffyOchs2009}; \citealt{DalBo.et.al2010}; \citealt{HarstadLanciaRusso2019}), or whenever the players' preferences are non-standard e.g. altruistic preferences (\citealt{Andreoni1990}; \citealt{FehrGatcher2000}; \citealt{FehrGatcher2002}). In a recent study, \citet{gallice2019co} (G\&M, henceforth) in the framework of a public good game, show that it is possible  to achieve full cooperation, even in the case where the interaction is one-shot, the number of players is finite, and the agents are self-interested. In this model, individuals have to make decisions sequentially, without knowing their position in the sequence (position uncertainty), but are aware of the decisions of some of their predecessors by observing a sample of past play. In the presence of position certainty, those placed in the early positions of the sequence would want to contribute, in an effort to induce some of the other group members to cooperate \citep{rapoport1994provision}, while late players, would want to free-ride on the contributions of the early players. Nevertheless, if the agents are unaware of their position in the sequence, they would base their choice on the average payoff, from all potential positions, and would be inclined to contribute  to induce potential successors to do the same. G\&M show that full contribution can occur in equilibrium, where given appropriate values of the parameters of the game (i.e. return from contributions), there exists an equilibrium where all agents contribute. G\&M extended these results to the case of a multi-player sequential prisoner's dilemma, showing that cooperation can be achieved when positional uncertainty is present. 

In this study we design and conduct an economic experiment where we aim to test the theoretical predictions of the G\&M model in a multi-player prisoner's dilemma environment, where subjects are not aware of their position in the sequence and receive only partial information on past cooperating actions. Our objective for this study is three-fold, as we aim to understand: $(1)$ Whether agents cooperate or defect under positional uncertainty, given the observed actions of their two immediate predecessors; $(2)$ How agents behave in case of positional awareness; $(3)$ Are there any other behavioural attributes which are exhibited, like altruism and social preferences \citep{charness2002}(C\&R, henceforth). To answer these questions, we vary the positional awareness of the subjects, and ask them to make conditional decisions in all potential information conditions about the predecessor(s) actions.  Subjects in groups of 5 participate in a sequential prisoner's dilemma game where some members of the group face positional uncertainty. Using the \textit{strategy method} elicitation method (\citealp{Selten1967}) we collect data from our participants varying the available information of past play. This allows us to test the predictions of the G\&M model and to also explore the presence of other behavioural types such as those with social preferences (modelled assuming C\&R  preferences), free riders or unconditional altruists. Using finite mixture modelling,  we are able to classify our experimental population to one of these types. We find that the majority of the subjects behave in a free-riding way, around 30\%  behave in line with the G\&M predictions, 10\% are subjects with altruistic tendencies and the remaining engage in either altruistic behaviour, or are types that cannot be clearly identified.  We conclude that the assumption of full rationality is not sufficient to lead to full cooperation and argue how the alternative behavioural models of social preferences are able to contribute to the interpretations of the data. 

Furthermore, we test the robustness of our design by introducing a within-subjects condition, where the choices of the subjects in the same game were elicited using the direct method. Our results extend  the \citet{brandts2000hot} design to its sequential case, and confirm that there is no difference in the inference between the two elicitation methods.

The rest of the paper is organised as follows. In the next section, we discuss the related literature and the contribution of our study. In section \ref{sec:theory} we present the main theoretical predictions of \citet{gallice2019co}. Section \ref{sec:design} presents the experimental design and the procedures, while section \ref{sec:results} presents the descriptive statistics and results of the structural econometric analysis. We then conclude.

\section{Related literature}
Prior to the contribution of G\&M, the timing of contributions in a public good and the total amount of funds raised had been the subject of significant theoretical investigation. \citet{Varian1994} demonstrates that, under specific assumptions, a sequential contribution mechanism can lead to lower total contributions than a simultaneous one. This result stems from the strategic structure of the model, in which a first mover may exploit a first-mover advantage to free-ride. In contrast, \citet{cartwright2010}, using a sequential public goods game with exogenous ordering, show that agents positioned early in the sequence may be willing to contribute if they anticipate that others will imitate their behaviour. In the context of fundraising, \citet{romano2001charities} examine when a charity might prefer to implement a sequential-move mechanism for announcing contributions, while \citet{vesterlund2003informational} finds that revealing past contributions can help signal an organization's quality and simultaneously reduce the free-rider problem faced by the fundraiser. \citet{AnwarBrunoFoucartSenGupta2025} extend G\&M to a setting where groups contribute simultaneously but act sequentially based on partial observations of past contributions. They show that groups conditionally cooperate to influence subsequent ones, making each agent pivotal in sustaining contributions. In the context of self-enforcing climate agreements—typically lacking enforcement for defection (\citealt{Barrett2003}; \citealt{Nordhaus2015})—their results suggest that group membership alone may sustain cooperation without formal enforcement.

In the framework of a prisoner's dilemma, previously, \citet{nishihara1997} analyse a $n$-person sequential prisoner's dilemma game with position uncertainty and shows how a planner can induce cooperation by immediately informing all agents when defection occurs. On the equilibrium path, agents cooperate for the same reason as they do in the G\&M model, i.e. to encourage potential successors towards cooperation. Off the equilibrium path, agents in \citet{nishihara1997} are immediately informed if a predecessor defects before their move, and hence, no agent can convince his successors to cooperate. Instead, G\&M focus on situations where observability is local and unconditional, and therefore an agent who observes defection may play a pivotal role in preventing further defection by choosing to cooperate.

Our study contributes to the literature of social preferences explaining decisions in sequential prisoner's dilemmas.  Prisoner's dilemma  have been thoroughly used  in the literature to test the presence of conditional cooperation (see for example \citealt{clark2001}, \citealt{blanco2011} for some early examples, and more recently in \citealt{Miettinen20},  \citealt{baader2024}, and \citealt{kirchsteiger2024}). Table \ref{tab:litreview} summarizes some recent papers examining cooperation in sequential prisoner's dilemma game. The various studies differ on a number of characteristics such as number of rounds, varying payoffs, provision of feedback, as well as on a number of more fundamental elements such as the way choices are elicited, either using the strategy (cold)  or the direct (hot) method, whether beliefs are elicited, whether there is an estimation of a structural model and finally, whether there is a classification of subjects to types. The majority of the studies use two-player games (only two papers in this list including ours are using larger groups), do not elicit beliefs, and do not estimate a structural model.  The closest studies to ours include \citet{vyrastekova2018},  \citet{baader2024} and \citet{Anwar2024}  where our results extend and complement theirs. 

\citet{vyrastekova2018} design and conduct an economic experiment to test the theoretical predictions of \citet{nishihara1997}. In a within-subjects experiment with 5-member groups, they study behaviour in two treatments, one with full and another without any information on the subjects' position in the sequences and subjects receive information on whether someone before them defected or not. Their design differs to ours in a number of aspects. First, in order for the \citet{nishihara1997} cooperative equilibrium to hold, they need to focus on a special class of social dilemmas where the payoff function must be convex in cooperation and concave in defection. In our design, using an alternative information mechanism, we are able to study behaviour in a multi-player sequential prisoner's dilemma , without the need to impose any constraints to the payoff function, extending their results to a more general case. Second, we elicit behaviour using both the direct and strategy method, and we also use structural estimations to classify subjects to various types. \citet{baader2024}  examine how the incentive to defect in a sequential prisoner's dilemma  affects conditional cooperation. Using 2-player games and eliciting behaviour from a large number of sequential prisoner's dilemmas with varying payoffs, they structurally estimate a model of social preferences, highlighting the role of positive weights on the opponent's payoff in conditional cooperation. In our study, we focus on a multi-person sequential prisoner's dilemmas and show how  their results extend to the case of positional uncertainty.  \citet{Anwar2024} developed an experiment  to test the predictions of G\&M in the \textit{public goods} game. They find that approximately 25\% of the subjects behaved in line with the theoretical predictions, while the majority were classified as conditional cooperators. Some subjects exhibited altruistic tendencies, while only a small minority engaged in free-riding behaviour. Using the same information mechanism, we extend these results to the case of sequential prisoner's dilemmas, showing that while the proportion of G\&M decision makers remains relatively constant, the majority of the players now adopt a free-riding approach.

From a methodological perspective, our study contributes to the literature comparing the strategy (cold) and direct (hot) elicitation methods (see, for example, \citealt{brandts2000hot}, \citealt{falk2005}, \citealt{reuben2012} in the context of sequential prisoner’s dilemma games). Prior research has found that these two methods generally produce indistinguishable behaviour, suggesting that eliciting a subject’s full strategy across all contingencies does not necessarily alter their choices during the experiment. Our results extend this finding to a more complex decision environment involving multiple information conditions and varying positions within the sequence across rounds. We find no significant differences in cooperation rates between the two elicitation methods, reinforcing the conclusions of \citet{brandts2011strategy}.

Finally, we also contribute to the finite mixture modelling literature. Most of the studies listed in Table \ref{tab:litreview} that classify subject to different types, usually base this classification on deterministic choice and a relatively number of elicitation questions (usually two). This approach, despite its efficiency, does not allow the researcher to investigate further types, and also is susceptible to noise, as potential trembles of the decision maker will be wrongly classified as a certain behavioural type. On the other hand, finite mixture modelling allows for both the identification of a number of discrete data generating processes, along with the estimation of error (stochasticity) in decision making. This econometric method has been extensively used in the literature of risky choice (see \citealt{bruhin10}; \citealt{conte11} among others) and has been recently extended in the literature of social preferences (see for example \citealt{bardsley2007experimetrics}; \citealt{iriberri2013elicited}; \citealt{conte2014econometric}; \citealt{bruhin19})

\begin{landscape}
\begin{table}[H]
\centering
\renewcommand{\arraystretch}{0.95}
\scriptsize
\begin{tabular}{lcclcccccl}
\hline
Study                         & Size of group & Rounds & Method   & Feedback & Beliefs & Treatments & Structural model & Types & Which types                                                                                                                               \\ \hline
\citet{ahn2007}              & 2             & 4      & strategy & no       & no      & 3          & no               & yes   & advantaged, disadvantaged, egoistic, conditional cooperator                                                                               \\
\citet{aksoy2014}      & 2             & 8*     & both     & no       & no      & 1          & yes              & no    & social value orientation model                                                                                                            \\
\citet{baader2024}         & 2             & 64*    & strategy & no       & yes     & -          & yes              & yes    & conditional cooperator, free-rider, unconditional cooperator, mismatcher                                                                                                                                         \\
\citet{bell2017}            & 2             & 2      & direct   & no       & no      & 3          & no               & yes   & positive and negative reciprocity, cooperation bias.                                                                                    \\
\citet{blanco2011}           & 2             & 1      & strategy & no       & no      & -          & yes              & no    & -                                                                                                                                         \\
\citet{blanco2014}         & 2             & 1      & strategy & no       & yes     & 3          & no               & no    & -                                                                                                                                         \\
\citet{brandts2000hot}    & 2             & 1      & both     & no       & no      & 2          & no               & no    & -                                                                                                                                         \\
\citet{clark2001}     & 2             & 10     & direct   & yes      & no      & 3          & no               & no    & -                                                                                                                                         \\
\citet{dhaene10}  & 2             & 1      & direct   & no       & yes     & -          & yes              & yes   & whether a material and//or a   reciprocity best response was given;                                                                       \\
\citet{eichenseer20}   & 2             & 1      & strategy & no       & no      & -          & no               & yes   & conditional cooperator; selfish; altruist; mismatcher                                                                                     \\
\citet{guzman2020}        & 2             & 60*    & strategy & yes      & no      & -          & yes              & no    &                                                                                                                                           \\
\citet{kirchsteiger2024}     & 2             & 30     & both     & yes      & no      & 4          & no               & yes   & conditional cooperator; selfish; altruist; mismatcher                                                                                     \\
\citet{Miettinen20}       & 2             & 1      & strategy & yes      & yes     & -          & no               & yes   & \begin{tabular}[c]{@{}l@{}}homo economicus, inequity aversion, conditional welfare, \\ reciprocity,   altruism, homo moralis\end{tabular} \\
\citet{ridinger2021}              & 2             & 20     & direct   & yes      & no      & 4          & no               & no    & -                                                                                                                                         \\
\citet{vyrastekova2018} & 5             & 10     & direct   & yes      & no      & 2          & no               & no    & -                                                                                                                                         \\
this study                    & 5             & 10     & both     & yes      & no      & 2          & yes              & yes   & free rider, altruist, social preferences, G\&M                                                                                            \\ \hline
\end{tabular}
\caption{Comparison of related studies examining cooperation in sequential prisoner's dilemma games. \textit{Method} refers to the elicitation approach, either direct response, strategy method, or both. \textit{Feedback} indicates whether subjects received feedback after each round. \textit{Beliefs} indicates whether participants were asked to state their beliefs about others' choices. \textit{Structural} refers to whether a structural econometric model was estimated, and \textit{Types} indicates whether subjects were classified into behavioural types. *Denotes studies in which each round involved a different set of payoffs.}
\label{tab:litreview}
\end{table}
\end{landscape}

\section{Theoretical Framework}\label{sec:theory}
In this section we present the main features and predictions of the model -- following the notation of  \citet{gallice2019co} -- and refer interested readers to the original study for further details. Consider a set $I=\{1, \ldots, n\}$ of risk-neutral agents making choices in a sequential prisoner's dilemma. They know the length $n$ of the sequence, but are uncertain of their position in the sequence. Players are exogenously placed in the sequence that determines the order of play, and they can be at any position with equal chances. Thus, each player has symmetric beliefs about her position in the sequence. When the turn of player \(i\in I\) arrives she can observe a \textit{sample} of her
predecessors’ action. She then choose one of two actions, \(a_i \in \{C,D\}\), where \(a_i=C\) indicates cooperation and \(a_i=D\) indicates defection. 
 Table \ref{tab:payoffs} shows the payoffs for the agents from a one-shot interaction (i.e. when matched with only one person in the sequence). 
\begin{table}[H]
\centering

\begin{tabular}{lcccll}
 & \multicolumn{1}{l}{} & \multicolumn{2}{c}{Player $j$} &  &  \\ \cline{2-4}
\multicolumn{1}{l|}{} & \multicolumn{1}{c|}{} & \multicolumn{1}{c|}{C} & \multicolumn{1}{c|}{D} &  &  \\ \cline{2-4}
\multicolumn{1}{c|}{\multirow{2}{*}{Player $i$}} & \multicolumn{1}{c|}{C} & \multicolumn{1}{c|}{1,1} & \multicolumn{1}{c|}{\textit{-l},1+g} &  &  \\ \cline{2-4}
\multicolumn{1}{c|}{} & \multicolumn{1}{c|}{D} & \multicolumn{1}{c|}{1+g,\textit{-l}} & \multicolumn{1}{c|}{0,0} &  &  \\ \cline{2-4}
 & \multicolumn{1}{l}{} & \multicolumn{1}{l}{} & \multicolumn{1}{l}{} &  & 
\end{tabular}
\caption{A Prisoner's Dilemma Game}
\label{tab:payoffs}
\end{table}

With $g>0$ and $l>0$, the game constitutes a prisoner’s dilemma. Agents sequentially choose whether to cooperate or defect. They receive  information about the actions of their $m$ immediate predecessors. After all players have chosen their actions, each agent is matched to all of her $n-1$ opponents in a series of pairwise prisoner's dilemma interactions. Agent $i$'s total payoff is the sum of the payoffs from each pairwise interaction. The payoff of the agent in case she cooperates is given by:
\begin{equation}\label{eq:utilityC}
    u_i(C,G_{-i})=G_{-i}-(n-1-G_{-i})l
\end{equation}
with $G_{-i}$ the number of players other than $i$ who cooperated. In case she defects, the payoff is given by:
\begin{equation}\label{eq:utilityD}
    u_i(D,G_{-i})=(1+g)G_{-i}
\end{equation}

We now introduce the notion of \textit{sampling}. Before taking an action, each player observes how many of her \( m\geq 2 \) predecessors have contributed. Players in positions 1 through \( m \) will observe fewer than \( m \) actions, given that there are less than \(m\) players before them . For example, if \( m = 2 \), the first-position player sees an empty sample, the second-position player sees a sample consisting of just one past action, and the third-position player observes a full sample of two past actions. Each sample is represented by the pair
\[
\xi = (\xi', \xi'')
\]
where the first element indicates the number of agents sampled, and the second element records the number of agents who cooperated within that sample.\footnote{For instance, \(\xi=(2,2)\) indicates that both of the previous 2 players cooperated; \(\xi=(2,0)\) indicates that neither of them cooperated; and \(\xi=(2,1)\) shows that only one of the two cooperated. Note that the first player in the sequence always receives the sample \(\xi=(0,0)\).} Importantly, a player cannot distinguish individual actions — only the overall sample size and the total number of cooperation. As a consequence, the first \( m \) agents can deduce their positional order based on the sample they observe. In this framework, players engage in an extensive form game with incomplete information, and G\&M uses the solution concept of \textit{sequential equilibrium} \citep{kreps1982}.

Let $\Xi$ be the set of all samples that a player observes. Player $i's$ strategy is a function \[\sigma_i(C\mid \xi): \Xi \to [0,1]\] that specifies the probability of cooperating given the sample received. Let $\sigma = \{\sigma_i\}_{i\in I}$ denote a strategy profile and $\mu = \{\mu_i\}_{i\in I}$ a system of beliefs. For a pair $(\sigma^*, \mu^*)$ to be a \textit{sequential equilibrium}, $\sigma^*$ should be (sequentially) rational given $\mu^*$, and $\mu^*$ should be consistent with $\sigma^*$. Let \(\Xi^C\) be the set of all samples without defection. Given this, \citet[Lemma~5, p.2147]{gallice2019co}  states that, assuming the following profile of play for all \(i \in I \):

\begin{align*}
    \sigma_i^*(C\mid \xi) =
\begin{cases}
1, & \text{if $\xi \in \Xi^C$ } \\
0, & \text{otherwise} 
\end{cases}.
\end{align*}

it follows that $(\sigma^*,\mu^*)$ is a sequential equilibrium provided that:
\begin{equation}\label{eq:condition1}
    g\leq 1-\frac{2m}{n+m-1}.
\end{equation}

In other words, a player will always cooperate when she observes a sample of size $m$ with full cooperation, otherwise she will defect.  Therefore, full cooperation emerges when agents play sequentially, are uncertain about their position, and observe the cooperating actions of some of their immediate $m$ predecessors. 
\begin{proof}
Consider first an agent who observes a sample with defection:  $\xi \notin \Xi^C$. An agent who observes defection never believes she can prevent further defection and therefore, defects herself. Now consider an agent who observes a sample of full cooperation $\xi \in \Xi^C$. According to the equilibrium definition, this occurs at the equilibrium path, and the agent can therefore infer that all the previous agents in the sequence cooperated. By cooperating herself, she knows that all subsequent players will cooperate as well, and therefore, her expected payoff from cooperating will be equal to $n-1$, and this payoff is independent of the agent's beliefs regarding her position in the sequence. On the other hand, the payoff from defecting depends on these beliefs. Consider an agent who observes a sample of $m$. This agent can therefore infer that she is not placed in the first $m$ positions, and there are equal chances of her being in any of the positions in $\{m+1,\cdots, n\}$. Let \(Q(i)\) denote agent's position. The expected position is:
\[E_{\mu}[Q(i)|\xi]=\frac{1}{n-m}\sum_{t=m+1}^nt=\frac{n+m+1}{2},\]
which means that she can expect that $(n+m-1)/2$ agents have already cooperated. Observing full cooperation (and therefore, assuming that everyone  beforehand has cooperated), the agent cooperates whenever:
\[E_{\mu^*}[u_i(\alpha_i,G_{-i})|\xi]\geq E_{\tilde{\mu}}[u_i(\alpha_i,G_{-i})|\xi],\]

\begin{equation}
  n-1\geq \frac{1+g}{n-m}\sum_{t=m+1}^n(t-1)=\frac{n+m-1}{2}(1+g).  \nonumber
\end{equation}

Therefore, the agent cooperates for values of $g$ that satisfy:

\begin{equation}\label{condition1b}
    g\leq \frac{n-m-1}{n+m-1}. \nonumber
\end{equation}

If the sample contains $\xi'<m$ total actions, the agent knows that she is in position $\xi'+1$ with certainty. The number of agents who have cooperated so far is $\xi'$ and the expected payoff from defecting is even lower ($\xi'$). The previous expression guarantees that agents in the first $m$ positions cooperate.
 \end{proof}

For further generality, suppose that mutual cooperation yields the reward payoff \(R\), defection against a cooperator yields the temptation payoff \(T\), cooperation against a defector yields the sucker payoff \(S\), and mutual defection yields the punishment payoff \(P\), the game can be written as follows:

\begin{table}[H]
\centering
\begin{tabular}{lcccll}
 & \multicolumn{1}{l}{} & \multicolumn{2}{c}{Player $j$} &  &  \\ \cline{2-4}
\multicolumn{1}{l|}{} & \multicolumn{1}{c|}{} & \multicolumn{1}{c|}{C} & \multicolumn{1}{c|}{D} &  &  \\ \cline{2-4}
\multicolumn{1}{c|}{\multirow{2}{*}{Player $i$}} & \multicolumn{1}{c|}{C} & \multicolumn{1}{c|}{\(R,R\)} & \multicolumn{1}{c|}{\(S,T\)} &  &  \\ \cline{2-4}
\multicolumn{1}{c|}{} & \multicolumn{1}{c|}{D} & \multicolumn{1}{c|}{\(T,S\)} & \multicolumn{1}{c|}{\(P,P\)} &  &  \\ \cline{2-4}
 & \multicolumn{1}{l}{} & \multicolumn{1}{l}{} & \multicolumn{1}{l}{} &  & 
\end{tabular}
\label{tab:general}
\caption{General form}
\end{table}
with the payoffs satisfying the conditions \footnote{Although $2R>T+S$ condition is primarily necessary in the repeated Prisoner’s Dilemma—where it ensures that alternating between cooperation (C) and defection (D) is not a profitable strategy—we retain it here to facilitate comparison with the standard approach.} \begin{equation} \label{eq:condition2}
    T>R>P>S
\end{equation} and 
\begin{equation}
    2R>T+S. \label{eq:condition3}
\end{equation} In this general structure an agent cooperates for values of  \(T\) that satisfy:

\begin{equation}\label{eq: econdition4}
    T\leq \frac{2(n-1)R-(n-m-1)P}{m+n-1}
\end{equation}

\begin{proof}
    
Replacing for a generic payoff structure with $R$ for cooperation and $T$ for temptation respectively, the utility from cooperating is equal to $(n-1)R$, while the utility from defection is: 
\begin{equation}
  \frac{n+m-1}{2}T+\frac{n-m-1}{2}P \nonumber
\end{equation}
as the player expects that $(n+m-1)/2$ players have cooperated and the remaining will defect after they observe defection. Therefore, to observe full cooperation, we need: 
\begin{equation}
 (n-1)R\geq  \frac{n+m-1}{2}T+\frac{n-m-1} {2}P \nonumber
\end{equation}
which holds whenever:

\begin{equation}
    T\leq \frac{2(n-1)R-(n-m-1)P}{m+n-1}.
\end{equation}
 \end{proof}

To keep it simple and relevant to our experiment, let us assume $n=5$ players in the group, and a sample of $m=2$. The model predicts: if players at positions 3, 4 and 5 in the sequence observe full cooperation, they will cooperate. If they observe at least one defection, the player believes that she cannot prevent further defections and, therefore, defects herself. Players in positions 1 and 2 (position certainty) have stronger incentives to cooperate. 

Appendix \ref{sec:appendix_gm} presents the analytical derivations of the G\&M model's predictions across all positions and various information conditions, which we summarize in Table \ref{tab:decisions_gm}. For each position and corresponding information condition ($m_c$) — that is, cooperation by both previous players, cooperation by only one, or defection by both — Table \ref{tab:decisions_gm} summarizes the expected utility (i.e., the expected payoffs) for each action. The table provides these utilities for a general group size ($n$) and sample size ($m$), as well as for the specific parameters used in our experiment ($n=5$, $m=2$). The last column reports the predicted choice of the decision maker, according to the G\&M model,  under the assumption that the constraints in the payoff matrix above hold.

\begin{table}[H]
\resizebox{\textwidth}{!}{%
\begin{tabular}{@{}ccccccc@{}}
\toprule
Position        & Info(\(m_c\)) & $EU_C$                                                            & $EU_D$                                                            & $EU_C$  & $EU_D$  & Decision \\ \midrule
\textgreater{}2 & 2    & $(n-1)R$                                                          & $\left(\frac{n+m-1}{2}\right)T+\left(\frac{n-m-1}{2}\right)P$     & $4R$    & $3T+P$  & C        \\
\textgreater{}2 & 1    & $\left(\frac{n+m-1}{2}-1\right)R+\left(\frac{n-m-1}{2}+1\right)S$ & $\left(\frac{n+m-1}{2}-1\right)T+\left(\frac{n-m-1}{2}+1\right)P$ & $2R+2S$ & $2T+2P$ & D        \\
\textgreater{}2 & 0    & $\left(\frac{n+m-1}{2}\right)S+\left(\frac{n-m-1}{2}\right)S$     & $\left(\frac{n+m-1}{2}\right)P+\left(\frac{n-m-1}{2}\right)P$     & $4S$    & $4P$    & D        \\
2               & 1    & $(n-1)R$                                                          & $T+(n-2)P$                                                       & $4R$    & $T+3P$  & C        \\
2               & 0    & $S+(n-2)S$                                                        & $P+(n-2)P$                                                        & $4S$    & $4P$    & D        \\
1               & -    & $(n-1)R$                                                          & $(n-1)P$                                                         & $4R$    & $4P$    & C        \\ \bottomrule
\end{tabular}%
}
\caption{The Table summarises the Expected Utility (expected payoff) for each action at each position in the sequence along with the corresponding predicted choice. The third and fourth columns reports these values for any size of the group ($n$) and and sample size ($m$), while the next two column report the same information for the case of a group of size 5 ($n=5$), where each player observes the choices of the two immediate predecessors ($m=2$). The last column reports the predicted choice assuming \citet{gallice2019co} preferences.  }
\label{tab:decisions_gm}
\end{table}
This allows us to state our first two hypotheses that we aim to test:
\begin{hyp}\label{hyp:hyp1}
There will be full cooperation when there is position uncertainty, and  the agents observe samples \(m\geq2\) of full cooperation. 
\end{hyp}
and a complementary one

\begin{hyp} \label{hyp:hyp2}
    cooperation will unravel when agents observe partial or no cooperation in samples of size $m\geq 2$.
\end{hyp}

\section{Experimental Design}\label{sec:design}
To test the predictions of the model presented in the previous section, we designed and conducted an incentivised economic experiment.  The experiment took place at the Lancaster Experimental Economics Lab (LExEL) in February 2025, involving {85} subjects (48.2 females) across four sessions and were mainly undergraduate students from various disciplines.\footnote{The size of the sample is based on a power analysis, conducted with a significance level of 0.05 and aiming for 80\% power, to detect a moderate effect size (Cohen's w = 0.2) in the specified McNemar’s test for paired binary data. The analysis indicated a minimum sample of 64 subjects and was conducted using the \textit{pwr.2p.test} function from the library \textit{pwr} in R.  For the power, we also used simulations for our structural model reported in section \ref{sec:results}. In these simulation we assumed a sample of 50 subjects which was sufficient for our estimation routine to identify the behavioural parameters of our statistical model. We report details of the simulation in Appendix \ref{sec:simulation}.}. Recruitment took place  via the ORSEE system \citep{greiner2015subject}, and the experiment was computerised using a custom-developed software written in Python\footnote{\citet{python}}. We excluded subjects who have participated in a prisoner's dilemma experiment in the past, and all subjects participated only once.

Each session was  divided into three parts. In Parts 1 and 3 participants play the game introduced earlier for 10 rounds each. We  elicited behaviour using both the \textit{strategy method} \citep{Selten1967} in Part 1  and the  \textit{direct method} in Part 3.  In Part 2, participants completed a real effort task. Scope of this part was to reduce cross-contamination between Parts 1 and 3. The order of the parts was the same for all participants. This fixed ordering was implemented to minimize potential anchoring effects, as our primary objective was to observe behaviour across all information conditions. Participants were informed that the experiment consisted of three parts, but the instructions for each part were provided only after the preceding part was completed.\footnote{The full set of instructions and screenshots from the experimental interface are provided in the Supplementary Material.}

In our experimental implementation we use the following payoffs which satisfy all the conditions in \eqref{eq:condition2}, \eqref{eq:condition3} and \eqref{eq: econdition4}, at an exchange rate of £1  per 250 tokens\footnote{The utilisation of both Experimental Currency Units (ECU) and the Random Payment Mechanism (RPM) has been a subject of ongoing debate in the literature. \citet{drichoutis2015veil} show that there is no difference between using ECUs or cash in the lab; while at the same time, the use of ECUs leads to decisions closer to theoretical predictions. Conversely, empirical research comparing the different incentive mechanisms is inconclusive \citep{azrieli2018incentives}, while \citet{azrieli2020incentives} provide a theoretical justification that the RPM is incentive-compatible in almost all experiments.
}.
\begin{table}[H]
\centering
\begin{tabular}{lcccll}
 & \multicolumn{1}{l}{} & \multicolumn{2}{c}{Player $j$} &  &  \\ \cline{2-4}
\multicolumn{1}{l|}{} & \multicolumn{1}{c|}{} & \multicolumn{1}{c|}{C} & \multicolumn{1}{c|}{D} &  &  \\ \cline{2-4}
\multicolumn{1}{c|}{\multirow{2}{*}{Player $i$}} & \multicolumn{1}{c|}{C} & \multicolumn{1}{c|}{500,500} & \multicolumn{1}{c|}{50,600} &  &  \\ \cline{2-4}
\multicolumn{1}{c|}{} & \multicolumn{1}{c|}{D} & \multicolumn{1}{c|}{600,50} & \multicolumn{1}{c|}{100,100} &  &  \\ \cline{2-4}
 & \multicolumn{1}{l}{} & \multicolumn{1}{l}{} & \multicolumn{1}{l}{} &  & 
\end{tabular}
\caption{Payoff Table}
\label{tab:game_matrix}
\end{table}

In each session, participants formed groups of 5 and played for a total of 10 rounds in both parts. To replicate the one-shot play environment, we  used a random matching protocol where in every round, participants were randomly matched into new groups.  At the beginning of all the sessions, participants were  given written instructions followed by a comprehension test. Participants were  able to retake the questionnaire until they pass. The interface  provided immediate feedback and explanation on the questions the subjects fail to respond to correctly. An experimenter was in the room to offer additional clarifications if there were further questions. Therefore, no exclusion criteria were applied. In every session, participants  played the sequential prisoner's dilemma game, as described in the previous section. In every round, subjects were randomly positioned in the sequence (with equal chances of being allocated at each position) and were sequentially asked to choose either  \(X\) (cooperation) or \(Y\) (defect). A participant would not learn about their position in the sequence unless they were at positions 1 or 2.  

To rigorously test the predictions of the model, we need information on what subjects’ choices are, in all potential samples they can observe. In particular, the model predicts that if a player observes full cooperation by the two previous players, she will cooperate herself, otherwise, if she observes at least one defection action, she will defect. In every round, participants were given different scenarios and were  asked to choose either \(X\) (cooperate) or \(Y\) (defect) for each of these scenarios. The scenarios presented to the participant  depended on their position. In particular, depending on their position,  participants were given at least one of these scenarios:
\begin{itemize}
    \item If the participant was at position 1 (i.e., there is no previous member), they would be asked to choose between \(X\) and \(Y\).
\item If the participant was at position 2, they would be asked to indicate what they would choose if the previous member chose \(X\) and what they would choose if the previous member chose \(Y\).
\item If the participant was at positions 3, 4, or 5, they would learn that their position is unknown. They would only be provided with information on the choices of the two immediate members before them. For instance, if the participant was at position 4, they would receive scenarios based on the choices of the members in positions 2 and 3, and so on. However, they would never find out the position of these members (and therefore in which position they themselves were). The participant would be asked to indicate what they would choose if the two previous members both chose \(X\), both chose \(Y\), and if one chose \(X\) and the other chose \(Y\).
\end{itemize}

Once all subjects submitted their choices, the payoffs for a subject were calculated according to the four pairwise matches which were based on the actual realisation of play (i.e. based on what player 1 chose, the software would retrieve the contingent choice of player 2 and so on). At the end of every round, feedback was  provided regarding the choices of the other members of the group, along with the corresponding payoffs. Subjects had access to an aggregate payoff table that informed them on their total payoffs that each of the available actions could generate, for all the potential choices of the other participants (e.g. if everyone defects, if only one person in the group defects and the other three cooperate, and so on)  We implemented a random payment mechanism paying one randomly selected round from each part. The average payment was £15.9, including a show up fee of £3. The sessions lasted less than 45 minutes and payments were made via bank-transfer. 

After Part 1 is completed participants were given instructions for Part 2. In Part 2 subjects were asked to complete a real effort task and in particular, they  were  asked to count the number of zeros in 5 randomly generated matrices (for a discussion on real effort tasks see \citealt{charness2018}). This was a filler task to reduce potential cross-contamination between Part 1 and Part 3. After Part 2 was completed participants received instructions for Part 3. Part 3 was identical to Part 1, but now the choices were elicited using the \textit{direct method}. 

The main goal of our experiment is to test the theoretical  predictions of G\&M and using the strategy method to elicit the full strategy of a player was crucial for our experiment, as it allows us to observe behaviour in all the potential information conditions (i.e., when both previous players cooperate, when only one cooperates, and when both defect). This means participants were asked to make contingent decisions for each possible information set depending on their randomly assigned position in the sequence.\footnote{The strategy method has been extensively used in the framework of prisoner's dilemma games (see \citealt{aksoy2014}; \citealt{brandts2000hot}; \citealt{kirchsteiger2024}) or public good games (see \citealt{herrmann2009measuring}; \citealt{fischbacher2010}; \citealt{teyssier2012inequity};  \citealt{martinsson2013}; \citealt{katuscak2023drives}), while previous studies found no statistical differences in subjects' responses between the strategy and the direct response method (see \citealt{brandts2000hot};  \citealt{brandts2011strategy}, or \citealt{keser2021strategy} for a discussion).} 
The secondary goal of our experiment is to examine whether the elicitation method influences participants' choices. Specifically, we aim to explore potential differences in decisions between those elicited using the strategy method and those using the direct method. \citet{brandts2000hot} were the first to test the direct-response method
versus the strategy method in a 2-player sequential Prisoner’s Dilemma\footnote{\citet{falk2005} and \citet{reuben2012} are comparing the two methods using a prisoner's dilemma game, but they do so in a different context. \citet{falk2005} studied a prisoner's dilemma game with sanctioning opportunities, finding mixed results on the effect of the two methods, while \citet{reuben2012} studied a repeated sequential prisoner's dilemma finding no significant difference. }. With direct-response, a first mover made a choice that was observed by the responder,
who then chose a response. With the strategy method, the responder had to make two contingent choices, one for each possible first-mover choice. We extend this framework to our multi-player sequential prisoner's dilemma.  This leads us to our third hypothesis:
\begin{hyp}\label{hyp:hyp3}
    There is no significant difference in behaviour between the \textit{hot} elicitation environment (\textit{direct}) and the \textit{cold} environment (\textit{strategy method}). 
\end{hyp}

\section{Results}\label{sec:results}
\subsection{Descriptive Statistics}
Before estimating our structural model, we first present some summary information on choices and cooperation rates. We focus on the data from Part 1 where choices are elicited using the strategy method and therefore, we have data on the full strategy profile of our subjects. More specifically, we have data from three potential information conditions, namely, the decisions made by subjects when they observe a sample (the choice of the 2 immediate predecessors) of only defection ($c_0$), the decisions made by the subjects when they observe one cooperation and one defection action ($c_1)$, and decisions made by subjects when they observe a sample of full cooperation ($c_2$). 

\begin{figure}[h!]
    \centering
    \includegraphics[width=0.75\linewidth]{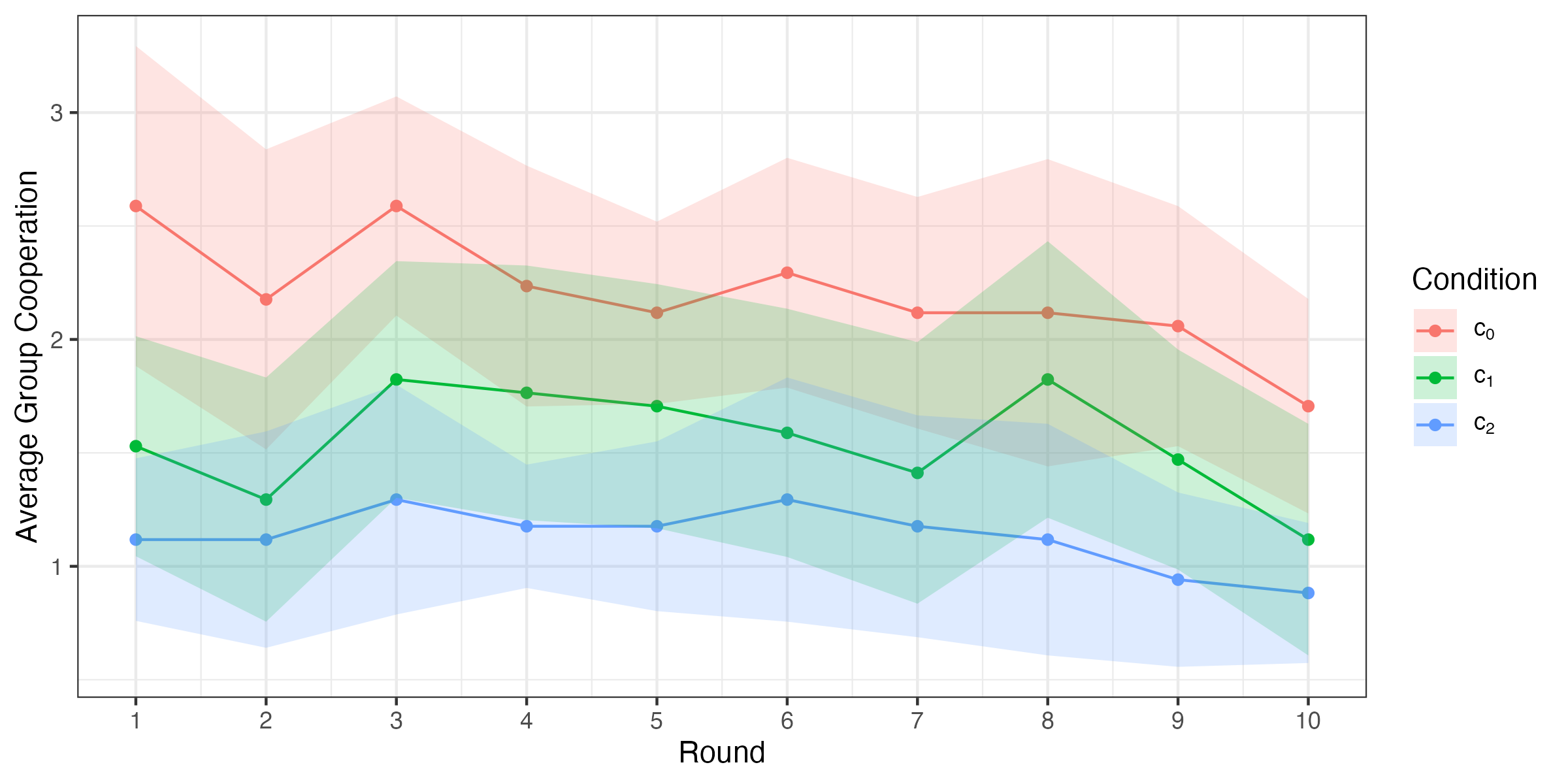}
    \caption{Group level cooperation across all conditions. The ribbon shows the 95\% confidence interval.}
    \label{fig:group_level}
\end{figure}

\begin{figure}[h!]
    \centering
    \includegraphics[width=0.75\linewidth]{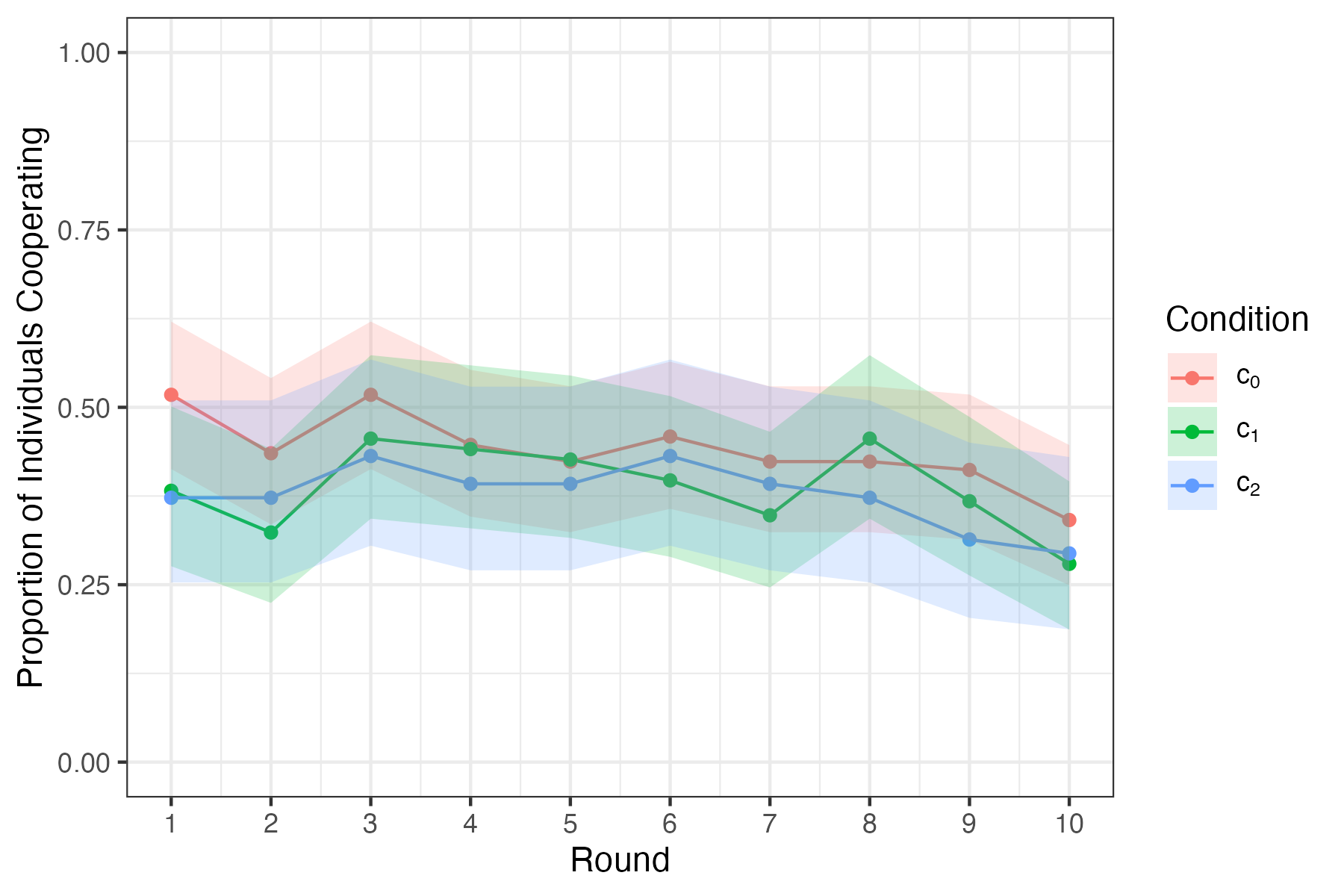}
    \caption{Individual level cooperation across all conditions. The ribbon shows the 95\% confidence interval.}
    \label{fig:individual_level}
\end{figure}

Figures~\ref{fig:group_level} and \ref{fig:individual_level} present aggregate cooperation rates over time at the group and individual levels, respectively. Contributions remain relatively stable across all rounds, with no significant differences between the first and last five rounds, and no evidence of end-of-period effects. Furthermore, the overlap in the confidence intervals suggests that the dynamics of cooperative outcomes are not significantly affected by the information condition.
Across all rounds, group-level cooperation is highest in condition $c_0$ (44\%), followed by $c_1$ (38.8\%) and $c_2$ (37.6\%). Notably, cooperation in condition $c_1$ exceeds that in $c_2$, running counter to the theoretical prediction that full cooperation observed in $c_2$ should encourage further cooperation.

 Let $\bar{c}_i$ denote the proportions of cooperation in conditions \(i \in \{0,1,2\}\). The theoretical prediction is that \( \bar{c}_2>\bar{c}_1> \bar{c}_0 \), but we instead observe the opposite effect \( \bar{c}_2<\bar{c}_1<\bar{c}_0 \).  One-sided exact binomial tests of $\bar{c}_2>\bar{c}_0 ~(p=0.9084)$ and $\bar{c}_1>\bar{c}_0 ~(p=0.9352)$ show no significant difference in the rate of cooperation between the two conditions\footnote{A pairwise McNemar's test shows that there is no statistically significant difference between the choices under the \(c_0\) and \(c_1\) conditions $(p = 0.1686)$, and also \(c_2\) and \(c_0\) $(p = 0.234)$.}. 

\begin{figure}[ht]
    \centering
    \begin{subfigure}[b]{0.45\textwidth}
        \includegraphics[width=\textwidth]{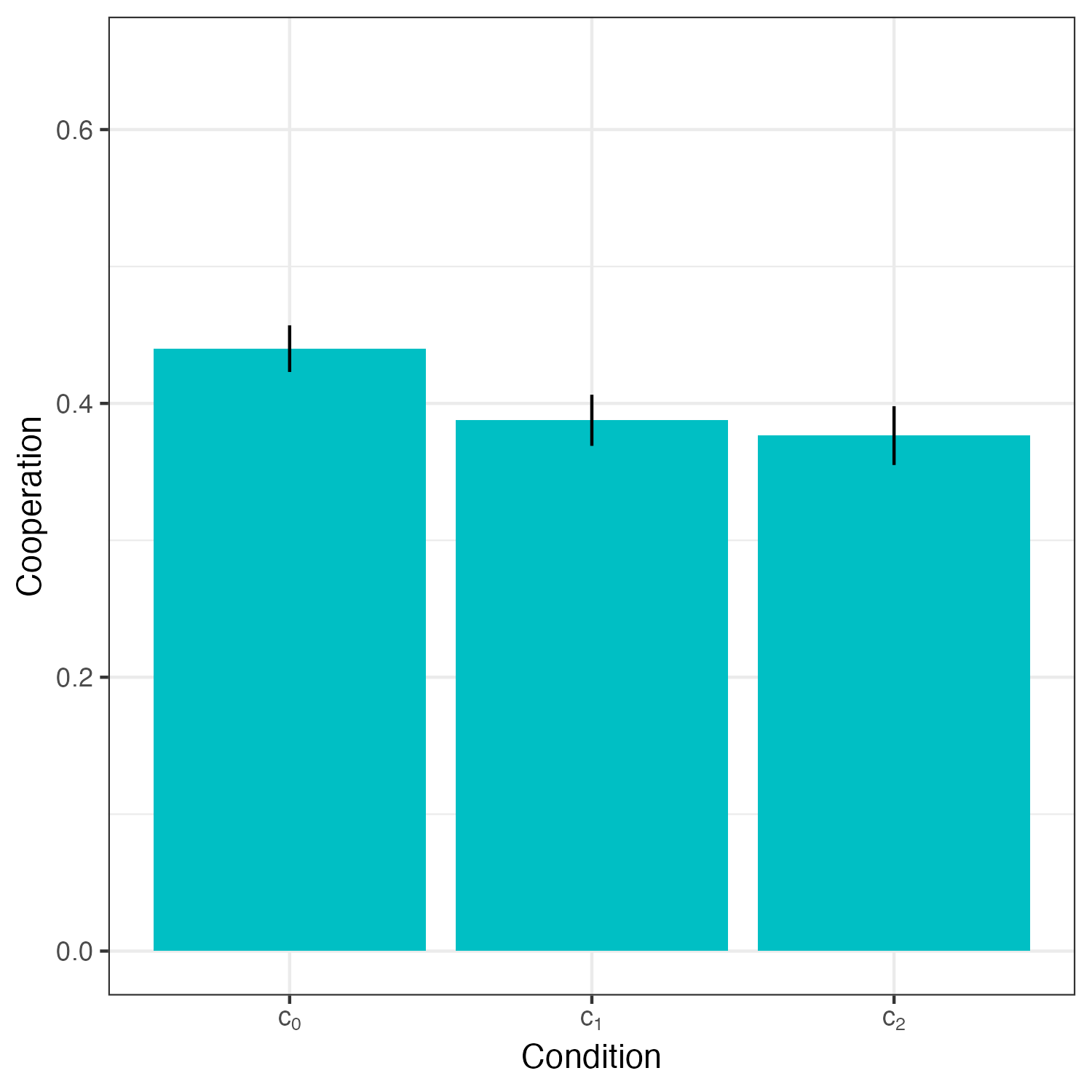}
        \caption{Overall Cooperation (All Positions)}
        \label{fig:plot_all}
    \end{subfigure}
    \hfill
    \begin{subfigure}[b]{0.45\textwidth}
        \includegraphics[width=\textwidth]{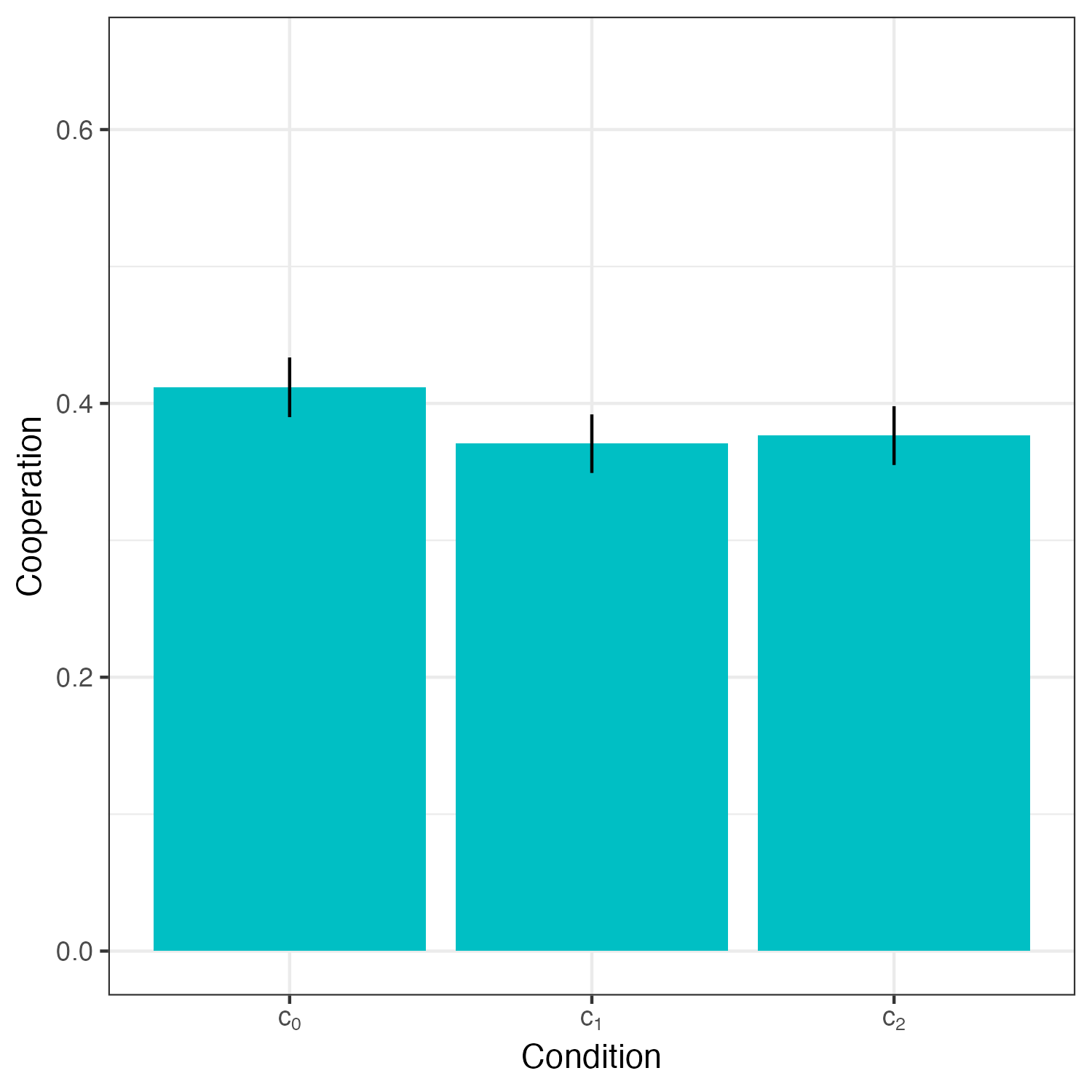}
        \caption{Cooperation (Positions 3-5)}
        \label{fig:plot_345}
    \end{subfigure}
    \caption{Comparison of Cooperation}
    \label{fig:comparison}
\end{figure}

Figure \ref{fig:comparison} illustrates the rate of cooperation in Part 1. The left panel shows the decisions of all the subjects in all possible positions while the right panel shows the decisions of subjects positioned 3rd, 4th or 5th. where they face uncertainty. In both panels cooperation level is higher in condition $c_0$ compared to condition $c_1$ and $c_2$. In the left panel, it appears that cooperation is decreasing in the condition. Nevertheless this difference is not statistically significant. 

At the individual level, the same pattern holds: cooperation in $c_0$ is consistently higher than in $c_1$ and $c_2$, except for round 8, where the trend temporarily reverses. These findings do not support the hypothesis that cooperation would be highest in condition $c_2$, relative to $c_0$ and $c_1$. This leads us to our first result:

\begin{fnd} \label{finding:1}
   There is no significant difference in cooperation levels when subjects observe full cooperation, partial cooperation, or no cooperation in their sample. There is a significant level of cooperation even in the conditions where subjects observe defection in the sample.
\end{fnd}

These results provide no support for the hypothesis that position uncertainty promotes cooperation, nor for the notion that cooperation unravels when defection occurs. This result is in line with \citet{vyrastekova2018} who find extensive free-riding and no difference between the no and full information treatments. 

Table~\ref{tab:rates} reports the cooperation rates both in aggregate and disaggregated by position within the sequence. Interestingly, cooperation rates tend to decline as a subject's position in the sequence increases, suggesting that later players are less likely to contribute. The rate of cooperation of those placed at position 1 is 52.9\% compared to 43.5\% of those placed in the second position, and 41.2\% of those we face position uncertainty, in no cooperation history condition. The difference between first and last places is highly significant (p=0.0097, McNemar's test). A similar pattern is observed in the other two conditions.

\begin{table}[H]
\centering
\begin{tabular}{@{}cccc@{}}
\toprule
Position & $c_0$ & $c_1$ & $c_2$ \\ \midrule
1 & 0.529 & - & - \\
2 & 0.435 & 0.441 & - \\
\textgreater{}2 & 0.412 & 0.370 & 0.441 \\
All & 0.440 & 0.388 & 0.376 \\ \bottomrule
\end{tabular}
\caption{Cooperation rates across all positions in the sequence.}
\label{tab:rates}
\end{table}

\begin{fnd} \label{finding:2}
   The rate of cooperation is declining as a function of the position of the subject in the sequence.
\end{fnd}
This finding is closely related to to the leading-by-example literature, based on the linear public goods game (see \citealt{gachter2012}; \citealt{levati2007leading}; \citealt{potters2007leading}; \citealt{guth2007leading}; \citealt{figuieres2012vanishing}; \citealt{sutter2014leadership}; \citealt{preget2016}, in the context of public goods games, or \citealt{moxnes2003effect}, in a public bad experiment). They all find robust evidence of first movers contributing more than later movers, and later movers' contributions to be positively correlated to the first movers' contributions, indicating reciprocal motives.

\subsection{Structural model}
While the results in the previous section do not provide support on the hypothesis that positional uncertainty would achieve full cooperation, there is a significant level of cooperation which appears to be independent of the information condition.  In this section, we use structural econometrics to investigate why subjects contribute and under which conditions. More specifically we use structural modelling  to distinguish between subjects who are best described by the G\&M model and subjects whose preferences display  alternative behavioural strategies (e.g. social preferences, conditional cooperation). Following the terminology adopted in the type-classification literature in public goods games (see for instance \citealt{fischbacher2001people}; \citealt{bardsley2007experimetrics}; \citealt{preget2016}; \citealt{thoni2018conditional};  \citealt{Miettinen20}; \citealt{katuscak2023drives} ) we define and explore the existence of the following types in our data: the \textit{G\&M} type, the \textit{free-rider} type, the \textit{altruist} type and the \textit{conditional cooperator} type. The G\&M type  behaves as presented in section \ref{sec:theory}. The free-rider and altruist types represent two extremes: the former never cooperates, while the latter always cooperates, no matter their position in the sequence. An intermediate case is reflected in the conditional cooperator type, who always cooperates if she is in position 1 and also cooperates if she is at any other position and at least another player in the sequence has cooperated. This type has a conditional concern for welfare which we model by assuming C\&R preferences (see related papers \citealt{Miettinen20}; \citealt{baader2024}, \citealt{bruhin19}). Following C\&R we assume that the utility of the players depends on both their own and others' earnings. In particular, letting $\pi_A$ and $\pi_B$ the  monetary payoffs for players A and B respectively, the utility of Player B  is given by the piecewise linear function:

\begin{align}
     U_B(\pi_A,\pi_B)
    \begin{cases} 
        (1-\rho)\pi_B+\rho \pi_A & \text{if }  \pi_B> \pi_A ~ \text{~(advantageous inequality)}\\
        (1-\sigma)\pi_B+\sigma \pi_A & \text{if }  \pi_B< \pi_A ~ \text{~(disadvantageous inequality)}.
    \end{cases}
\end{align}

where $\sigma$ and $\rho$ parameters that capture various aspects of social preferences. This formulation says that the utility of Player B, is a weighted sum of her own monetary and the others player's payoff and may depend on whether the other player is getting  a higher or lower payoff. Depending on the values of $\sigma$ and $\rho$, subjects belong to different preference types. When both are zero, the model reduces to a purely selfish type and the predictions of the model coincide with those of G\&M.  When $\sigma<0$ the subject is behindness averse, where she weights the other player's payoff negatively whenever her payoff is smaller than the other's. When $\rho>0$, the subject is aheadness averse, that is she weights the payoff of the other player positively, whenever her payoff is larger than the other player. Consequently, a subject with $\sigma <0 < \rho$ is both behindness and aheadness aversion, and furthermore, if $-\sigma<\rho$ then this subject is characterised as a difference averse type for whom disadvantageous inequality bears less weight than advantageous inequality. Similarly, in the case where $\sigma <0 < \rho$ and $\sigma>\rho$, the subject is still difference averse but now disadvantageous inequality is more important than advantageous inequality. Preferences with both $\sigma$ and $\rho$ positive characterise an altruistic type, while negative values of the parameters indicate spiteful behaviour. 

Our main objective is to model the behaviour of the \textit{conditional cooperator} type, the one who cooperates when she observes cooperation from at least one past predecessor, and defects otherwise.  In Appendix \ref{sec:appendix_cr} we provide the choice predictions of this model in the framework of our experiment, for three specifications. 
\subsection{Estimation method}
To estimate the proportion of our experimental population that is best characterised by each type we adopt a finite mixture modelling approach  (see 
\citealt{mclachlan2000finite}, for an overview) which is based on the random utility framework \citep{McFadden1981}. Mixture models have been widely used to estimate models of risky choice (see, for example, \citealt{bruhin10}; \citealt{conte11}), and have more recently been applied in the social preferences literature to identify other-regarding types of behaviour \citep{bruhin19}\footnote{An alternative methodology is the  Strategy Frequency Estimation Method (SFEM) which was introduced by \citet{dalbo11} and \citet{fudenberg12} to estimate the distribution of behavioural types in experimental data. It assumes individuals follow one of a finite set of deterministic strategies and incorporates a fixed probability of implementation error (a tremble), independent of payoffs. SFEM has since been widely applied in the study of cooperation games, including \citet{bigoni2015time}, \citet{breitmoser2015}, \citet{frechette2017infinitely}, \citet{dalbo2019} and \citet{Anwar2024}. A related approach is proposed by \citet{bardsley2007experimetrics} in the context of continuous public goods games. In contrast, finite mixture models based on the random utility framework assume that individuals evaluate choices according to expected utility and that observed choices reflect stochastic utility differences across options. The likelihood of choosing a given option increases with the utility difference between the options. See \citet{mclachlan2000finite} for a comprehensive overview of finite mixture models, and \citet[p.~3930]{dalbo2019} for a discussion of alternative strategy estimation methods and related identification issues.}.

A finite mixture model is flexible enough to identify the existence of distinct preference types, allowing for heterogeneity in decision making within the population. Mixture models assume that the population is made up by a finite number of $\mathcal{K}$ distinct preference types, each characterised by its own set of parameters $\theta$.  Consider a subject $i \in \{1,2,\dots,N\}$ who is best described by type $k$ in the set of types $\mathcal{K}=\{gm, coop, alt, free\}$. Let $\pi_k$ the population share of type $k$, with $\sum_{k=1}^K\pi_k=1$. In round $r$ the observed choice of individual $i$ is  a binary choice denoted by $y_i^r\in\{C,D\}$ for \textit{cooperate} and \textit{defect} respectively. 
There are in total four types, two of which are utility based (both G\&M and C\&R assume maximisation of expected utility), and two of which are heuristic-based (both the free-rider and altruist type always adopt the same strategy\footnote{These preferences can also be modelled  assuming \citet{charness2002} preferences, when the parameters of the model are constrained in a particular range. Nevertheless, doing so would cause identification issues}). To model stochasticity in decision making, we apply \citet{McFadden1981} random utility model for discrete choices augmented with a tremble term, modelling both structured preferences and random errors. Specifically, we assume that choices are governed by  a logit link function over deterministic utilities to capture stochasticity in preference evaluation. To account for additional behavioural noise such as inattention or response slips, we incorporate a tremble parameter $\omega$\footnote{See \citet{moffatt2001} for a discussion on how omitting to account for a tremble can lead to biased coefficients, and \citet{conte11} for an implementation in a mixture model.}. The resulting probability of choosing an option is a convex combination between the model predicted choice and uniform random choice:
\begin{equation}
    P(y_i^r=C;\theta_k,\beta,\omega)=     (1-\omega)\frac{exp(\beta EU_c)}{exp(\beta EU_c)+exp(\beta EU_d)}+\frac{\omega}{2}
\end{equation}
where $EU_c$ and $EU_d$ are the corresponding expected utilities of cooperation and defection as defined in section \ref{sec:theory} and Appendix \ref{sec:appendix_cr},  $\theta_k$ the vector of preference parameters of model, $k$, $\omega\in (0,1/2)$ the tremble parameter, and  $\beta$ a parameter capturing the choice sensitivity with respect to the deterministic difference between the expected utility of the two options. When $\beta \rightarrow 0$ the model predicts random choice, and when $\beta \rightarrow  \infty$ the probability of choosing the option with the highest expected utility converges to 1. 

For the two types that  assume heuristic decision making, we assume that the decision maker behaves according to the constant error model (see \citealt{harless94}), where with probability $1-\omega$ she chooses the action that the heuristic prescribes, and with probability $\omega$ she chooses the opposite \footnote{For parsimony reasons and to reduce the total number of parameters in the statistical model,  we assume that all types share the same parameters for $\beta$ and $\omega$. Relaxing this hypothesis does not alter the results qualitatively.}. The likelihood for individual $i$ is given by:
\begin{equation}
\mathcal{L}=\sum_{k=1}^K\pi_k\prod^R_{r=1}P(y^r_i|\theta_k,\beta,\omega)
\end{equation}
where each individual's likelihood is a mixture of type-specific probabilities. The full sample log-likelihood is therefore given by:
\begin{equation}
    log\mathcal{L}=\sum_{i=1}^N \log\left(\sum_{k=1}^K\pi_k\prod^R_{r=1}P(y^r_i|\theta_k,\beta,\omega)\right)
\end{equation}
Objective is to find optimal values for $\beta, \omega, \theta$,  and $\pi$ that maximise the overall likelihood across all subjects $I$ and set of types (strategies) $\mathcal{K}$. Therefore,there are seven parameters to estimate: two for the C\&R model ($\rho$ and $\sigma$), the precision and tremble parameters ($\beta$ and $\omega$), and the population shares for three types—$\pi_{gm}$ (G\&M), $\pi_{coop}$ (conditional cooperator), and $\pi_{free}$ (free-rider). The mixing probability $\pi_{alt}$ for the altruist is simply the residual probability.  We estimate the models using Maximum Likelihood Estimation techniques.\footnote{For the estimation we use a general nonlinear augmented Lagrange multiplier optimisation routine that allows for random initialisation of the starting parameters as well as multiple restarts of the solver, to avoid local maxima. The estimation was conducted using the \emph{R} programming language for statistical computing (The \emph{R} Manuals, version 4.3.1. Available at: http://www.r-project.org/). The estimation codes are available in the replication repository.}

 All three specifications differ on the way the conditional cooperator decision maker is modelled. We explore three different definitions, one which assumes a modified version of the G\&M model to account for social preferences, one that allows for the existence of a pure conditional cooperator, and one where we use the \citet{charness2002} multi-player setting along with the notion of the \emph{reciprocal fairness equilibrium}. We provide details on how each decision maker makes her optimal choices in Appendix \ref{sec:appendix_cr}.  Furthermore, in Appendix \ref{sec:simulation} we report the results of an extensive Monte Carlo simulation we performed to verify that our experimental design is appropriate to identify behavioural heterogeneity and that our econometric implementation is suitable to successfully estimate the parameters of our statistical model.

Table~\ref{tab:estimates} reports the parameter estimates ($\hat{\theta}, \hat{\pi}_k, \hat{\beta}, \hat{\omega}$) of the mixture model under the three different specifications. Across all models, the pattern of type classification remains broadly consistent. Based on the values of the Akaike and Bayesian Information Criteria,\footnote{The Akaike Information Criterion (AIC) and the Bayesian Information Criterion (BIC) are model selection tools that balance model fit and complexity. In our mixture model, they help determine which specification best explains the data without overfitting, with lower values indicating a better trade-off between goodness of fit and parsimony. They are defined as:
\[
\text{AIC} = -2 \cdot \log L + 2k, \quad \text{BIC} = -2 \cdot \log L + k \cdot \log(n)
\]
where $\log L$ is the log-likelihood of the model, \( k \) is the number of estimated parameters, and \( n \) is the number of observations.}
the specification that models the conditional cooperator as a modified version of the G\&M framework—accounting for social preferences—provides the best fit to the data. As such, the remainder of the discussion focuses on the estimates from this model (first column on the table).

The majority of participants are classified as free-riders, comprising 48\% of the sample. This finding aligns with \citet{baader2024} who find the majority of their subjects behaving in a free-riding way, (49.3\% on average varying between 42.6 and 57.0\%) and \citet{kirchsteiger2024}, who report similar proportions of unconditional defectors (ranging from 38.6\% to 54.1\% across four treatments), as well as with \citet{vyrastekova2018} and \citet{eichenseer20}, who also observe high, albeit non-majority, shares of such types. The G\&M type accounts for 27.6\% of the population, closely matching the 25\% found in a public goods context by \citet{Anwar2024}. Furthermore, 10\% of subjects are classified as conditional cooperators, and the remaining 14.5\% either exhibit altruistic behaviour or could not be clearly classified into a specific type. With regards to the conditional cooperator type, our estimates suggest that subjects place substantial positive weight on others’ outcomes when disadvantaged ($\sigma=2.377$), consistent with reciprocal or fairness-motivated cooperation. Interestingly, the estimated value of $\rho=-1.2$, implies negative weighting of others' payoffs when ahead, which has been interpreted in the literature (e.g., \citealp{bruhin19}) as implausible or indicative of spiteful or competitiveness preferences. The values of the parameters for this model predict that this decision maker always cooperates unless she is at a position with uncertainty and observes full contribution. In that case the prediction is that this agent will defect. This pattern of behaviour could partially explain why there is a drop in cooperation in the full cooperation condition. Interestingly, this prediction runs counter to standard conditional cooperation logic and highlights how the structure of social preferences—particularly a negative weight on others' payoffs when ahead—can generate strategic defection even in highly cooperative environments. This result is very much in line with \citet{bruhin19} result, who find that their behindness averse type comprises roughly 10\% of the population. 

\begin{fnd}
    Based on the estimates of our structural model, the majority (48\%) of our experimental population can be classified as free-riders, 27.6\% as G\&M decision makers, 10\% as C\&R, and the remaining 14.5\% as altruists or  individuals whose type could not be clearly identified. 
\end{fnd}

\renewcommand{\arraystretch}{0.85} 
\begin{table}[H]
\centering
\begin{tabular}{@{}cccc@{}}
\toprule
Parameter    & (1)      & (2)      & (3)      \\ \midrule
$\pi_{gm}$   & 0.276*** & 0.292*** & 0.330*** \\
s.e.         & (0.074) & (0.075) & (0.077) \\
$\pi_{coop}$ & 0.100**  & 0.088*  & 0.099    \\
s.e.         & (0.037) & (0.040) & (0.069) \\
$\pi_{free}$ & 0.480*** & 0.471*** & 0.457*** \\
s.e.         & (0.066) & (0.065) & (0.067) \\
$\pi_{alt}$  & 0.145     & 0.149    & 0.114    \\
s.e.         & (0.106) & (0.107) & (0.123) \\
$\sigma$     & 2.377*  & 4.734**  & -        \\
s.e.         & (1.030) & (1.828) & -        \\
$\rho$       & -1.219* & -0.477   & -        \\
s.e.         & (0.560) & (0.404) & -        \\
$\gamma$     & -        & -        & 1.000    \\
s.e.         & -        & -        & (0.677) \\
$\delta$     & -        & -        & 0.000    \\
s.e.         & -        & -        & (0.945) \\
$\beta$      & 0.623** & 0.521** & 0.282*    \\
s.e.         & (0.209) & (0.180) & (0.146) \\
$\omega$     & 0.195*** & 0.191*** & 0.186*** \\
s.e.         & (0.017) & (0.016) & (0.017) \\ \midrule
LL           & -1186.544 & -1191.416 & -1210.32 \\
AIC          & 2387.088 & 2396.832 & 2434.646 \\
BIC          & 2426.433 & 2436.177 & 2473.991 \\
Obs          & 2040     & 2040     & 2040     \\ \bottomrule
\end{tabular}
\caption{The table reports the MLE estimates. All models include the same types, except for the specification of the conditional cooperator. Model 1 incorporates a modified version of G\&M allowing for social preferences. Model 2 introduces social preferences motivation, while Model 3 includes a social welfare consideration. Standard errors are reported in parentheses. LL denotes the log-likelihood value, AIC the Akaike Information Criterion, and BIC the Bayesian Information Criterion. Significance levels: $^{***} p < 0.01$, $^{**} p < 0.05$, $^{*} p < 0.1$.
}
\label{tab:estimates}
\end{table}

\subsection{Strategy versus the direct-response method}
Finally, we test whether the elicitation method had any impact on the cooperation rate. \citet{brandts2000hot} were the first to test the direct-response method versus the strategy method in a 2-player sequential Prisoner’s Dilemma\footnote{\citet{falk2005} and \citet{reuben2012} are comparing the two methods using a prisoner's dilemma game, but they do so in a different context. \citet{falk2005} studied a prisoner's dilemma game with sanctioning opportunities, finding mixed results on the effect of the two methods, while \citet{reuben2012} studied a repeated sequential prisoner's dilemma finding no significant difference. }. With direct-response, a first mover made a choice that was observed by the responder, who then chose a response. With the strategy method, the responder had to make two contingent choices, one for each possible first-mover choice. We now compare the results from Part 1 and Part 3 to examine the impact of the two elicitation methods — the strategy method (“cold”) and the direct method (“hot”). Part 3 was a repetition of Part 1, but choices were now elicited using the direct method. In each round, groups were re-matched, and players made decisions in a truly sequential manner. As a result, while the data from Part 1 include a full strategy profile for each player (i.e., choices for each possible history), in Part 3 subjects made only one decision per round, based on their position in the sequence and the actions of the two preceding players. To enable a direct comparison between the two methods, we convert the conditional choices from Part 1 into actual choices by matching the observed histories from Part 3. Aggregating over all decisions, the frequency of choosing action C in Part 1 was 43.2\% (367/850), which is quite close to the 45.9\% (390/850) observed in Part 3. This difference is not significant based on a McNemar's test $(p=0.236)$.

\begin{figure}
    \centering
    \includegraphics[width=0.5\linewidth]{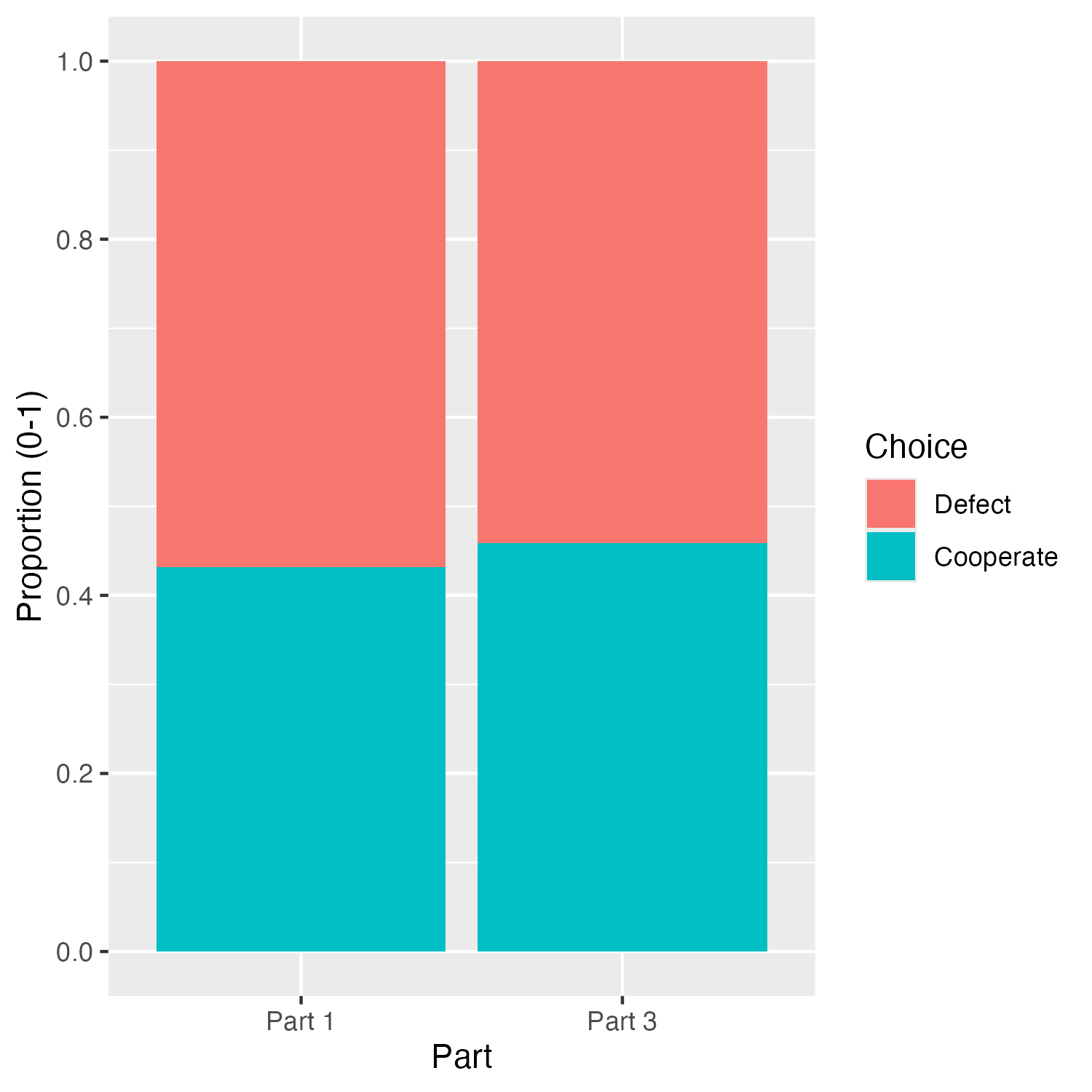}
    \caption{Hot vs Cold}
    \label{fig:Hot vs Cold}
\end{figure}

Figure \ref{fig:Hot vs Cold} presents the level of cooperation/defection in Part and Part 3. We can see there's not much difference between them. Using McNemar's test we find that there is no significant difference between elicitation methods \((p=0.2362)\), therefore we fail to reject Hypothesis \ref{hyp:hyp3}.

\begin{fnd}
    The elicitation method (whether decisions were made via the strategy method or direct method) does not significantly influence participants' decisions to cooperate or defect.
\end{fnd}

\section{Conclusion}
Understanding the conditions under which cooperation emerges in social dilemmas is essential, as such behaviour characterises the functioning of societies where individual and collective interests often conflict. While a range of mechanisms—such as punishment, reputation, communication, and institutional design—have been proposed in the literature to sustain cooperation, empirical evidence on their effectiveness remains context-dependent and incomplete. In other cases, cooperation appears to rely heavily on the presence of social preferences, such as altruism or reciprocity, which motivate individuals to act beyond narrow self-interest.

In a recent study, \citet{gallice2019co} show that full cooperation can be achieved in the framework of social dilemmas (public goods game and prisoner's dilemma), even when the interaction is one-shot, the number of players is finite, and individuals are purely self-interested. In their model, players make decisions sequentially without knowing their exact position in the sequence (i.e., under position uncertainty), but they do observe a sample of previous decisions made by others. The theoretical prediction is that players will cooperate when they observe full cooperation, otherwise they will defect. 

In this study, we design and conduct an economic experiment to test the theoretical predictions of \citet{gallice2019co}.  In a within-subject experiment with two elicitation conditions: the strategy method (cold) and the direct method (hot), subjects in groups of 5, participate in a sequential prisoner's dilemma game where they face positional uncertainty. Our findings indicate that the information structure does not significantly affect cooperation levels, consistent with \citet{vyrastekova2018}, who report similar results in a related game featuring a different information design. Nevertheless, we find a positive and significant level of cooperation in all information conditions, indicating the presence of alternative behavioural types. Using structural econometric analysis, we further classify subjects to those who are consistent with the predictions of the G\&M model, those who behave in a conditionally cooperating manner, modelled according to the predictions of the \citet{charness2002} model of social preferences, and to those who unconditionally cooperate (altruists) or unconditionally defect (free-riders). We find that the majority of the subjects resort to free-riding behaviour (48\%), around 27\% behave according to the predictions of G\&M, 10\% according to the predictions of  \citet{charness2002} while the remaining as best described as altruists. Finally, we find no difference in behaviour between the two elicitation methods, providing further support to the strategy elicitation method \citep{Selten1967} as a valid instrument to elicit behaviour in an interactive environment. 

Our results have implications both at a methodological and a policy making level. From a methodological point of view, the findings from the  literature on the stability and transferability of cooperation types across games remain mixed, with recent work aiming to understand the conditions under which cooperative tendencies generalise. \citet{eichenseer20} compare type classifications in a sequential Prisoner’s Dilemma  and a one-shot public goods game, and find that conditional cooperators are consistently identified across games. However, they caution that sequential prisoner's dilemmas are less reliable for identifying purely selfish types. Similarly, \citet{mullett2019} demonstrate significant but imperfect cross-task predictability between public good games and binary social dilemmas, concluding that prosociality behaves like a stable trait but is also context-dependent. 

In the context of our study, this debate is directly relevant. In the sequential Prisoner's Dilemma, we find that a majority of subjects are best described as free riders, with approximately 27\% aligning with Gallice and Monzon’s strategic model which also entail a large amount of defection behaviour. In contrast, \citet{Anwar2024}, using the same information mechanism in a public goods game—albeit with a different subject pool—found that altruists and conditional cooperators make up over 65\% of participants. These findings suggest that the nature of the strategic environment and institutional framing significantly affect the salience and detectability of social preference types, reinforcing conclusions by \citet{eichenseer20} that some types (e.g., conditional cooperators) are more consistently expressed across games, while others are more context-sensitive.

Our results may also have a direct impact on the modelling of policy negotiations and the design of effective interventions. Consider, for example, the case of international environmental policy (e.g., climate policy). Recent literature has debated the appropriate game-theoretic framework for capturing such interactions. \citet{decanio2013} provide a thorough analysis of how climate negotiations between two players can map onto different strategic environments depending on the underlying payoffs—such as the No-Conflict Game, Prisoner's Dilemma, Coordination Game, Chicken, Type Game, or Cycle. They conclude that under heightened perceptions of risk or long-term consequences, the structure of the interaction shifts away from a pure prisoner's dilemma dynamic and is better described as a coordination game—an argument further supported by \citet{mielke2018}. \citet{alessio2021} highlight the conditions under which this coordination may break down. Our results lend empirical support to the Climate Clubs proposal by \citet{Nordhaus2015}, who proposes the introduction of institutional mechanisms—such as selective incentives or trade sanctions—to shift the payoff structure and align incentives across heterogeneous actors. These clubs may be especially effective in stabilising cooperation among self-interested (free-riding) types while simultaneously enabling reciprocity-driven or prosocial countries to coordinate on cooperative outcomes. Future work could explore how institutional designs can be tailored to populations with differing strategic dispositions, and how behavioural insights from laboratory settings can inform the dynamics of real-world climate negotiations.

 \bibliography{references}{}

\begin{thebibliography}{}

\bibitem[Ahn et~al., 2007]{ahn2007}
Ahn, T.~K., Lee, M., Ruttan, L., and Walker, J. (2007).
\newblock Asymmetric payoffs in simultaneous and sequential prisoner's dilemma games.
\newblock {\em Public Choice}, 132(3--4):353--366.

\bibitem[Aksoy and Weesie, 2014]{aksoy2014}
Aksoy, O. and Weesie, J. (2014).
\newblock Hierarchical bayesian analysis of outcome- and process-based social preferences and beliefs in dictator games and sequential prisoner's dilemmas.
\newblock {\em Social Science Research}, 45:98--116.

\bibitem[Alessio and Manfredi, 2022]{alessio2021}
Alessio, M. and Manfredi, P. (2022).
\newblock Are climate games prisoners’ dilemmas?
\newblock {\em Ecological Economics}, 190:107211.

\bibitem[Andreoni, 1990]{Andreoni1990}
Andreoni, J. (1990).
\newblock Impure altruism and donations to public goods: a theory of warm glow giving.
\newblock {\em Economic Journal}, 100:464 -- 477.

\bibitem[Anwar et~al., 2025]{AnwarBrunoFoucartSenGupta2025}
Anwar, C. M.~S., Bruno, J., Foucart, R., and SenGupta, S. (2025).
\newblock Efficient public good provision between and within groups.
\newblock {\em Games and Economic Behaviour}, 150:183 -- 190.

\bibitem[Anwar and Georgalos, 2024]{Anwar2024}
Anwar, C. M.~S. and Georgalos, K. (2024).
\newblock Position uncertainty in a sequential public goods game: An experiment.
\newblock {\em Experimental Economics}, 27:820--853.

\bibitem[Azrieli et~al., 2018]{azrieli2018incentives}
Azrieli, Y., Chambers, C.~P., and Healy, P.~J. (2018).
\newblock Incentives in experiments: A theoretical analysis.
\newblock {\em Journal of Political Economy}, 126(4):1472--1503.

\bibitem[Azrieli et~al., 2020]{azrieli2020incentives}
Azrieli, Y., Chambers, C.~P., and Healy, P.~J. (2020).
\newblock Incentives in experiments with objective lotteries.
\newblock {\em Experimental Economics}, 23:1--29.

\bibitem[Baader et~al., 2024]{baader2024}
Baader, M., Gächter, S., Lee, K., and Sefton, M. (2024).
\newblock Social preferences and the variability of conditional cooperation.
\newblock {\em Economic Theory}.
\newblock Forthcoming, Open access.

\bibitem[Banerjee, 1992]{Banerjee1992}
Banerjee, A.~V. (1992).
\newblock A simple model of herd behavior.
\newblock {\em The Quarterly Journal of Economics}, 107:797--817.

\bibitem[Bardsley and Moffatt, 2007]{bardsley2007experimetrics}
Bardsley, N. and Moffatt, P. (2007).
\newblock The experimetrics of public goods: Inferring motivations from contributions.
\newblock {\em Theory and Decision}, 62(2):161--193.

\bibitem[Barrett, 1994]{Barrett1994}
Barrett, S. (1994).
\newblock Self-enforcing international environmental agreements.
\newblock {\em Oxford Economic Papers}, 46:878--894.

\bibitem[Barrett, 2003]{Barrett2003}
Barrett, S. (2003).
\newblock Environment and statecraft: the strategy of environmental treaty-making.
\newblock {\em OUP Oxford}.

\bibitem[Bell et~al., 2017]{bell2017}
Bell, R., Mieth, L., and Buchner, A. (2017).
\newblock Separating conditional and unconditional cooperation in a sequential prisoner's dilemma game.
\newblock {\em PLOS ONE}, 12(11):e0187952.

\bibitem[Bigoni et~al., 2015]{bigoni2015time}
Bigoni, M., Casari, M., Skrzypacz, A., and Spagnolo, G. (2015).
\newblock Time horizon and cooperation in continuous time.
\newblock {\em Econometrica}, 83(2):587--616.

\bibitem[Blanco et~al., 2014]{blanco2014}
Blanco, M., Engelmann, D., Koch, A.~K., and Normann, H.-T. (2014).
\newblock Preferences and beliefs in a sequential social dilemma: A within-subjects analysis.
\newblock {\em Games and Economic Behavior}, 87:122--135.

\bibitem[Blanco et~al., 2011]{blanco2011}
Blanco, M., Engelmann, D., and Normann, H.-T. (2011).
\newblock A within-subject analysis of other-regarding preferences.
\newblock {\em Games and Economic Behavior}, 72(2):321--338.

\bibitem[Brandts and Charness, 2000]{brandts2000hot}
Brandts, J. and Charness, G. (2000).
\newblock Hot vs. cold: sequential responses in simple experimental games.
\newblock {\em Experimental Economics}, 2:227--238.

\bibitem[Brandts and Charness, 2011]{brandts2011strategy}
Brandts, J. and Charness, G. (2011).
\newblock The strategy versus the direct-response method: a first survey of experimental comparisons.
\newblock {\em Experimental Economics}, 14:375--398.

\bibitem[Breitmoser, 2015]{breitmoser2015}
Breitmoser, Y. (2015).
\newblock Cooperation, but no reciprocity: Individual strategies in the repeated prisoner's dilemma.
\newblock {\em American Economic Review}, 105(9):2882--2910.

\bibitem[Bruhin et~al., 2019]{bruhin19}
Bruhin, A., Fehr, E., and Schunk, D. (2019).
\newblock The many faces of human sociality.
\newblock {\em Journal of the European Economic Association}, 17(4):1025--1069, 1335.

\bibitem[Bruhin et~al., 2010]{bruhin10}
Bruhin, A., Fehr-Duda, H., and Epper, T. (2010).
\newblock {Risk and Rationality: Uncovering Heterogeneity in Probability Distortion}.
\newblock {\em Econometrica}, 78(4):1375--1412.

\bibitem[Cartwright and Patel, 2010]{cartwright2010}
Cartwright, E. and Patel, A. (2010).
\newblock Imitation and the incentive to contribute early in a sequential public good game.
\newblock {\em Journal of Public Economic Theory}, 12(4):691--708.

\bibitem[Celen and Kariv, 2004]{CelenKariv2004}
Celen, B. and Kariv, S. (2004).
\newblock Observational learning under imperfect information.
\newblock {\em Games and Economic Behaviour}, 47:72 -- 86.

\bibitem[Charness et~al., 2018]{charness2018}
Charness, G., Gneezy, U., and Henderson, A. (2018).
\newblock Experimental methods: Measuring effort in economics experiments.
\newblock {\em Journal of Economic Behavior \& Organization}, 149:74--87.

\bibitem[Charness and Rabin, 2002]{charness2002}
Charness, G. and Rabin, M. (2002).
\newblock Understanding social preferences with simple tests.
\newblock {\em The Quarterly Journal of Economics}, 117(3):817--869.

\bibitem[Clark and Sefton, 2001]{clark2001}
Clark, K. and Sefton, M. (2001).
\newblock The sequential prisoner's dilemma: Evidence on reciprocation.
\newblock {\em The Economic Journal}, 111(468):51--68.

\bibitem[Conte et~al., 2011]{conte11}
Conte, A., Hey, J.~D., and Moffatt, P.~G. (2011).
\newblock Mixture models of choice under risk.
\newblock {\em Journal of Econometrics}, 162(1):79 -- 88.

\bibitem[Conte and Moffatt, 2014]{conte2014econometric}
Conte, A. and Moffatt, P. (2014).
\newblock The econometric modelling of social preferences.
\newblock {\em Theory and Decision}, 76(2):119--145.

\bibitem[Dal~Bo et~al., 2010]{DalBo.et.al2010}
Dal~Bo, P., Foster, A., and Putterman, L. (2010).
\newblock Institutions and behavior: experimental evidence on the effects of democracy.
\newblock {\em American Economic Review}, 100:2205 -- 2229.

\bibitem[Dal~B{\'o} and Fr{\'e}chette, 2019]{dalbo2019}
Dal~B{\'o}, P. and Fr{\'e}chette, G.~R. (2019).
\newblock Strategy choice in the infinitely repeated prisoner's dilemma.
\newblock {\em American Economic Review}, 109(11):3929--3952.

\bibitem[Dal~Bó and Fréchette, 2011]{dalbo11}
Dal~Bó, P. and Fréchette, G.~R. (2011).
\newblock The evolution of cooperation in infinitely repeated games: Experimental evidence.
\newblock {\em American Economic Review}, 101(1):411--29.

\bibitem[DeCanio and Fremstad, 2013]{decanio2013}
DeCanio, S.~J. and Fremstad, A. (2013).
\newblock Game theory and climate diplomacy.
\newblock {\em Ecological Economics}, 85:177--187.

\bibitem[Dhaene and Bouckaert, 2010]{dhaene10}
Dhaene, G. and Bouckaert, J. (2010).
\newblock Sequential reciprocity in two-player, two-stage games: An experimental analysis.
\newblock {\em Games and Economic Behavior}, 70(2):289--303.

\bibitem[Drichoutis et~al., 2015]{drichoutis2015veil}
Drichoutis, A.~C., Lusk, J.~L., and Nayga, R.~M. (2015).
\newblock The veil of experimental currency units in second price auctions.
\newblock {\em Journal of the Economic Science Association}, 1:182--196.

\bibitem[Duffy and Ochs, 2009]{DuffyOchs2009}
Duffy, J. and Ochs, J. (2009).
\newblock Cooperative behavior and the frequency of social interaction.
\newblock {\em Games and Economic Behaviour}, 66:785 -- 812.

\bibitem[Eichenseer and Moser, 2020]{eichenseer20}
Eichenseer, M. and Moser, J. (2020).
\newblock Conditional cooperation: Type stability across games.
\newblock {\em Economics Letters}, 188:108941.

\bibitem[Falk et~al., 2005]{falk2005}
Falk, A., Fehr, E., and Fischbacher, U. (2005).
\newblock Driving forces behind informal sanctions.
\newblock {\em Econometrica}, 73(6):2017--2030.

\bibitem[Fehr and Gatcher, 2000]{FehrGatcher2000}
Fehr, E. and Gatcher, S. (2000).
\newblock Cooperation and punishment in public goods experiments.
\newblock {\em American Economic Review}, 90:980 -- 994.

\bibitem[Fehr and Gatcher, 2002]{FehrGatcher2002}
Fehr, E. and Gatcher, S. (2002).
\newblock Altruistic punishments in humans.
\newblock {\em Nature}, 415 (6868):137.

\bibitem[Figuieres et~al., 2012]{figuieres2012vanishing}
Figuieres, C., Masclet, D., and Willinger, M. (2012).
\newblock Vanishing leadership and declining reciprocity in a sequential contribution experiment.
\newblock {\em Economic Inquiry}, 50(3):567--584.

\bibitem[Fischbacher and Gächter, 2010]{fischbacher2010}
Fischbacher, U. and Gächter, S. (2010).
\newblock Social preferences, beliefs, and the dynamics of free riding in public goods experiments.
\newblock {\em American Economic Review}, 100(1):541--56.

\bibitem[Fischbacher et~al., 2001]{fischbacher2001people}
Fischbacher, U., Gächter, S., and Fehr, E. (2001).
\newblock Are people conditionally cooperative? evidence from a public goods experiment.
\newblock {\em Economics Letters}, 71(3):397--404.

\bibitem[Fr{\'e}chette and Yuksel, 2017]{frechette2017infinitely}
Fr{\'e}chette, G.~R. and Yuksel, S. (2017).
\newblock Infinitely repeated games in the laboratory: four perspectives on discounting and random termination.
\newblock {\em Experimental Economics}, 20(2):279--308.

\bibitem[Friedman, 1971]{Friedman1971}
Friedman, J.~W. (1971).
\newblock A non-cooperative equilibirum of supergames.
\newblock {\em Review of Economic Studies}, 38:1 -- 12.

\bibitem[Fudenberg et~al., 2012]{fudenberg12}
Fudenberg, D., Rand, D.~G., and Dreber, A. (2012).
\newblock Slow to anger and fast to forgive: Cooperation in an uncertain world.
\newblock {\em American Economic Review}, 102(2):720--49.

\bibitem[Gallice and Monz{\'o}n, 2019]{gallice2019co}
Gallice, A. and Monz{\'o}n, I. (2019).
\newblock Co-operation in social dilemmas through position uncertainty.
\newblock {\em Economic Journal}, 129(621):2137--2154.

\bibitem[Greiner, 2015]{greiner2015subject}
Greiner, B. (2015).
\newblock Subject pool recruitment procedures: organizing experiments with orsee.
\newblock {\em Journal of the Economic Science Association}, 1(1):114--125.

\bibitem[Guzm{\'a}n et~al., 2020]{guzman2020}
Guzm{\'a}n, R., Harrison, R., Abarca, N., and Villena, M.~G. (2020).
\newblock A game-theoretic model of reciprocity and trust that incorporates personality traits.
\newblock {\em Journal of Behavioral and Experimental Economics}, 84:101497.

\bibitem[Gächter et~al., 2012]{gachter2012}
Gächter, S., Nosenzo, D., Renner, E., and Sefton, M. (2012).
\newblock Who makes a good leader? cooperativeness, optimism, and leading-by-example.
\newblock {\em Economic Inquiry}, 50(4):953--967.

\bibitem[Güth et~al., 2007]{guth2007leading}
Güth, W., Levati, V., Sutter, M., and van~der Heijden, E. (2007).
\newblock Leading by example with and without exclusion power in voluntary contribution experiments.
\newblock {\em Journal of Public Economics}, 91:1023--1042.

\bibitem[Harless and Camerer, 1994]{harless94}
Harless, D. and Camerer, C. (1994).
\newblock {The Predictive Utility of Generalised Expected Utility Theories}.
\newblock {\em Econometrica}, 62(6):1251--1289.

\bibitem[Harstad et~al., 2019]{HarstadLanciaRusso2019}
Harstad, B., Lancia, F., and Russo, A. (2019).
\newblock Compliance technology and self-enforcing agreements.
\newblock {\em Journal of European Economic Association}, 17:1 -- 29.

\bibitem[Herrmann and Thöni, 2009]{herrmann2009measuring}
Herrmann, B. and Thöni, C. (2009).
\newblock Measuring conditional cooperation: A replication study in russia.
\newblock {\em Experimental Economics}, 12(1):87--92.

\bibitem[Iriberri and Rey-Biel, 2013]{iriberri2013elicited}
Iriberri, N. and Rey-Biel, P. (2013).
\newblock Elicited beliefs and social information in modified dictator games: What do dictators believe other dictators do?
\newblock {\em Quantitative Economics}, 4(4):515--547.

\bibitem[Katuščák and Miklánek, 2023]{katuscak2023drives}
Katuščák, P. and Miklánek, T. (2023).
\newblock What drives conditional cooperation in public good games?
\newblock {\em Experimental Economics}, 26(2):435--467.

\bibitem[Keser and Kliemt, 2021]{keser2021strategy}
Keser, C. and Kliemt, H. (2021).
\newblock The strategy method as an instrument for the exploration of limited rationality in oligopoly game behavior (strategiemethode zur erforschung des eingeschränkt rationalen verhaltens im rahmen eines oligopolexperimentes).
\newblock In Charness, G. and Pingle, M., editors, {\em The Art of Experimental Economics: Twenty Top Papers Reviewed}, pages 30--37. Routledge, 1st edition.

\bibitem[Kirchsteiger et~al., 2024]{kirchsteiger2024}
Kirchsteiger, G., Lenaerts, T., and Suchon, R. (2024).
\newblock Voluntary versus mandatory information disclosure in the sequential prisoner’s dilemma.
\newblock {\em Economic Theory}.

\bibitem[Kreps and Wilson, 1982]{kreps1982}
Kreps, D.~M. and Wilson, R. (1982).
\newblock Sequential equilibria.
\newblock {\em Econometrica}, 50(4):863--894.

\bibitem[Levati et~al., 2007]{levati2007leading}
Levati, M.~V., Sutter, M., and Van~der Heijden, E. (2007).
\newblock Leading by example in a public goods experiment with heterogeneity and incomplete information.
\newblock {\em Journal of Conflict Resolution}, 51(5):793--818.

\bibitem[Martinsson et~al., 2013]{martinsson2013}
Martinsson, P., Pham-Khanh, N., and Villegas-Palacio, C. (2013).
\newblock Conditional cooperation and disclosure in developing countries.
\newblock {\em Journal of Economic Psychology}, 34:148--155.

\bibitem[McFadden, 1981]{McFadden1981}
McFadden, D. (1981).
\newblock Econometric models of probabilistic choice.
\newblock In Manski, C.~F. and McFadden, D., editors, {\em Structural Analysis of Discrete Data with Econometric Applications}, pages 198--272. MIT Press, Cambridge, MA.

\bibitem[McLachlan and Peel, 2000]{mclachlan2000finite}
McLachlan, G. and Peel, D. (2000).
\newblock {\em Finite Mixture Models}.
\newblock Wiley Series in Probability and Statistics. Wiley, New York.

\bibitem[Mielke and Steudle, 2018]{mielke2018}
Mielke, J. and Steudle, G.~A. (2018).
\newblock Strategies for climate cooperation: Can equity trump self-interest?
\newblock {\em Ecological Economics}, 150:250--260.

\bibitem[Miettinen et~al., 2020]{Miettinen20}
Miettinen, T., Kosfeld, M., Fehr, E., and Weibull, J. (2020).
\newblock Revealed preferences in a sequential prisoners’ dilemma: A horse-race between six utility functions.
\newblock {\em Journal of Economic Behavior \& Organization}, 173:1--25.

\bibitem[Moffatt and Peters, 2001]{moffatt2001}
Moffatt, P.~G. and Peters, S.~A. (2001).
\newblock Testing for the presence of a tremble in economic experiments.
\newblock {\em Experimental Economics}, 4(3):221--228.

\bibitem[Moxnes and Van~der Heijden, 2003]{moxnes2003effect}
Moxnes, E. and Van~der Heijden, E. (2003).
\newblock The effect of leadership in a public bad experiment.
\newblock {\em Journal of Conflict Resolution}, 47(6):773--795.

\bibitem[Mullett et~al., 2019]{mullett2019}
Mullett, T.~L., Brown, G.~D., and Tunney, R.~J. (2019).
\newblock Cooperation in public goods games predicts behavior in incentive‐matched binary dictator games.
\newblock {\em Economic Inquiry}, 57(3):1462--1476.

\bibitem[Nishihara, 1997]{nishihara1997}
Nishihara, K. (1997).
\newblock A resolution of n-person prisoners' dilemma.
\newblock {\em Economic Theory}, 10(3):531--540.

\bibitem[Nordhaus, 2015]{Nordhaus2015}
Nordhaus, W. (2015).
\newblock Climate clubs: overcoming free-riding in international climate policy.
\newblock {\em American Economic Review}, 105:1339 -- 1370.

\bibitem[Potters et~al., 2007]{potters2007leading}
Potters, J., Sefton, M., and Vesterlund, L. (2007).
\newblock Leading-by-example and signaling in voluntary contribution games: an experimental study.
\newblock {\em Economic Theory}, 33(1):169--182.

\bibitem[Pr{\'e}get et~al., 2016]{preget2016}
Pr{\'e}get, R., Nguyen-Van, P., and Willinger, M. (2016).
\newblock Who are the voluntary leaders? experimental evidence from a sequential contribution game.
\newblock {\em Theory and Decision}, 81(4):581--599.

\bibitem[{Python Software Foundation}, 2023]{python}
{Python Software Foundation} (2023).
\newblock Python language reference, version 3.10.
\newblock \url{https://www.python.org}.
\newblock Accessed: 2024-03-31.

\bibitem[Rapoport and Erev, 1994]{rapoport1994provision}
Rapoport, A. and Erev, I. (1994).
\newblock Provision of step-level public goods: Effects of different information structures.
\newblock In Schulz, U., Albers, W., and Mueller, U., editors, {\em Social Dilemmas and Cooperation}, pages 55--73. Springer-Verlag, New York.

\bibitem[Reuben and Suetens, 2012]{reuben2012}
Reuben, E. and Suetens, S. (2012).
\newblock Revisiting strategic versus non-strategic cooperation.
\newblock {\em Experimental Economics}, 15(1):24--43.

\bibitem[Ridinger, 2021]{ridinger2021}
Ridinger, G. (2021).
\newblock Intentions versus outcomes: Cooperation and fairness in a sequential prisoner's dilemma with nature.
\newblock {\em Games}, 12(3):58.

\bibitem[Romano and Yildirim, 2001]{romano2001charities}
Romano, R. and Yildirim, H. (2001).
\newblock Why charities announce donations: a positive perspective.
\newblock {\em Journal of Public Economics}, 81(3):423--447.

\bibitem[Selten, 1967]{Selten1967}
Selten, R. (1967).
\newblock {\em Die Strategiemethode zur Erforschung des eingeschränkt rationalen Verhaltens im Rahmen eines Oligopolexperiments}, pages 136--168.
\newblock Mohr, Tübingen.

\bibitem[Suleiman et~al., 1994]{suleiman1994position}
Suleiman, R., Budescu, D.~V., and Rapoport, A. (1994).
\newblock The position effect: The role of a player's serial position in a resource dilemma game.
\newblock In Schulz, U., Albers, W., and Mueller, U., editors, {\em Social Dilemmas and Cooperation}, pages 55--73. Springer-Verlag, New York.

\bibitem[Sutter and Rivas, 2014]{sutter2014leadership}
Sutter, M. and Rivas, M.~F. (2014).
\newblock Leadership, reward and punishment in sequential public goods experiments.
\newblock In Van~Lange, P. A.~M., Rockenbach, B., and Yamagishi, T., editors, {\em Reward and Punishment in Social Dilemmas}, pages 133--157. Oxford University Press.

\bibitem[Teyssier, 2012]{teyssier2012inequity}
Teyssier, S. (2012).
\newblock Inequity and risk aversion in sequential public good games.
\newblock {\em Public Choice}, 151(1):91--119.

\bibitem[Thöni and Volk, 2018]{thoni2018conditional}
Thöni, C. and Volk, S. (2018).
\newblock Conditional cooperation: Review and refinement.
\newblock {\em Economics Letters}, 171:37--40.

\bibitem[Varian, 1994]{Varian1994}
Varian, H.~R. (1994).
\newblock Sequential contributions to public goods.
\newblock {\em Journal of Public Economics}, 53(2):165--186.

\bibitem[Vesterlund, 2003]{vesterlund2003informational}
Vesterlund, L. (2003).
\newblock The informational value of sequential fundraising.
\newblock {\em Journal of Public Economics}, 87(3-4):627--657.

\bibitem[Vyrastekova and Funaki, 2018]{vyrastekova2018}
Vyrastekova, J. and Funaki, Y. (2018).
\newblock Cooperation in a sequential dilemma game: How much transparency is good for cooperation?
\newblock {\em Journal of Behavioral and Experimental Economics}, 77:88--95.

\end{thebibliography}
\bibliographystyle{apalike}

\newpage
\begin{appendices}
\section{Optimal choices in the \citet{gallice2019co} model}\label{sec:appendix_gm}
In this Appendix we present the optimal decisions according to the predictions of the \citet{gallice2019co} model for all potential  positions and all information sets a player may find herself during the experiment. We first present the solution for any size of group $n$ and then focus on the case of our experimental design where $n=5$. 
In our experimental implementation we use the following payoffs which satisfy  the following conditions:  $T>R>P>S$ and $2R>T+S$.
\begin{table}[H]
\centering
\begin{tabular}{lcccll}
 & \multicolumn{1}{l}{} & \multicolumn{2}{c}{Player $j$} &  &  \\ \cline{2-4}
\multicolumn{1}{l|}{} & \multicolumn{1}{c|}{} & \multicolumn{1}{c|}{C} & \multicolumn{1}{c|}{D} &  &  \\ \cline{2-4}
\multicolumn{1}{c|}{\multirow{2}{*}{Player $i$}} & \multicolumn{1}{c|}{C} & \multicolumn{1}{c|}{500,500} & \multicolumn{1}{c|}{50,600} &  &  \\ \cline{2-4}
\multicolumn{1}{c|}{} & \multicolumn{1}{c|}{D} & \multicolumn{1}{c|}{600,50} & \multicolumn{1}{c|}{100,100} &  &  \\ \cline{2-4}
 & \multicolumn{1}{l}{} & \multicolumn{1}{l}{} & \multicolumn{1}{l}{} &  & 
\end{tabular}
\end{table}
Let \(m=2\) denote the sample size, and let \(m_c\) denote the number of cooperators observed in the sample( so \(0 \leq m_c \leq m\)). We consider three potential positions in the sequence :  position 1 and 2, which entail decision making under certainty, and position >2, which involves positional uncertainty.  
\begin{description}

    \item[Position>2, \(m_c=2\).] 
    With a full sample of previous cooperating actions and position uncertainty, the player can infer that everyone before in the sequence has cooperated, and she also expects that playing $C$ will motivate the next player in the sequence to also play $C$, while playing D will lead to defection from the next player. The corresponding payoffs are given by:
    \begin{align*}
        U_c &= (n-1)R= 4R \\
        U_d &= \left(\frac{n+m-1}{2}\right)T+\left(\frac{n-m-1}{2}\right)P\\
        &=3T+P.
    \end{align*}
    where 
    \[\left(\frac{n+m-1}{2}\right)\]
is the expected position in the sequence and 

    \[\left(\frac{n-m-1}{2}\right)\]
is the number of the remaining participants in the sequence. Substituting for $m=2$ and $n=5$ we can show that the player always cooperates whenever $4R\geq 3T+P$. 

    \item[Position>2, \(m_c=1\).] 
     When observing only one prior cooperative action, according to the G\&M model, the player defects as she has no way to persuade the next player to cooperate. 
     The expected payoffs are given by:

    \begin{align*}
        U_c &= \left(\frac{n+m-1}{2}-1\right)R+\left(\frac{n-m-1}{2}+1\right)S\\
        &=2R+2S\\
        U_d &= \left(\frac{n+m-1}{2}-1\right)T+\left(\frac{n-m-1}{2}+1\right)P\\
        &=2T+2P.
    \end{align*}
    The player chooses the cooperating action whenever $R+S>T+P$ which given the structure of the payoffs in our experimental design, it can never happen. 
    \item[Position>2, \(m_c=0\).] 
    A player observing a sample of zero cooperating actions can infer that no-one in the sequence before her has cooperated and also she has no way to incentivise the subsequent player to do so. 
    \begin{align*}
    U_c&=\left(\frac{n+m-1}{2}\right)S+\left(\frac{n-m-1}{2}\right)S\\
    &=4S\\
    U_d&=\left(\frac{n+m-1}{2}\right)P+\left(\frac{n-m-1}{2}\right)P\\
    &=4P\end{align*}
    The player never cooperates since $P>S$ by design.
    \item[Position=2, \(m_c=1\)] 
    When the player is at position 2 and observes the  cooperating action from the first player, she can then trigger cooperation or defection for all the subsequent players based on her choice. If she cooperates, then all the subsequent players are expected to cooperate. If she defects, all subsequent players will defect. The expected payoffs are given by:
    \begin{align*}
    U_c&=(n-1)R=4R\\
    U_d&=T+(n-2)P=T+3P
    \end{align*}

    The player cooperates whenever $4R\geq T+3P$
    
    \item[Position=2, \(m_c=0\)] 
    A player observing a sample of zero cooperating actions can infer that the first player did not cooperate and also she has no way to incentivise the subsequent player to do so. 
    \begin{align*}
    U_c&=S+(n-2)S=4S\\
    U_d&=P+(n-2)P=4P
    \end{align*}
    The player always defects since $P>S$ by design. cooperates when the condition below holds:
     \item[Position=1]
     The first player in the sequence expects that if she cooperates (defects) everyone will cooperate (defect). The expected utility is given by:
         \begin{align*}
    U_c&=(n-1)R=4R\\
    U_d&=(n-1)P=4P
    \end{align*}
    which predicts that the player at the first position always contributes since $R>P$ by design. Table \ref{tab:decisions_gm} summarises the expected utility or each decision at each position and the prediction of the G\&M model.  
\end{description}

\begin{table}[H]
\resizebox{\textwidth}{!}{%
\begin{tabular}{@{}ccccccc@{}}
\toprule
Position        & Info(\(m_c\)) & $EU_C$                                                            & $EU_D$                                                            & $EU_C$  & $EU_D$  & Decision \\ \midrule
\textgreater{}2 & 2    & $(n-1)R$                                                          & $\left(\frac{n+m-1}{2}\right)T+\left(\frac{n-m-1}{2}\right)P$     & $4R$    & $3T+P$  & C        \\
\textgreater{}2 & 1    & $\left(\frac{n+m-1}{2}-1\right)R+\left(\frac{n-m-1}{2}+1\right)S$ & $\left(\frac{n+m-1}{2}-1\right)T+\left(\frac{n-m-1}{2}+1\right)P$ & $2R+2S$ & $2T+2P$ & D        \\
\textgreater{}2 & 0    & $\left(\frac{n+m-1}{2}\right)S+\left(\frac{n-m-1}{2}\right)S$     & $\left(\frac{n+m-1}{2}\right)P+\left(\frac{n-m-1}{2}\right)P$     & $4S$    & $4P$    & D        \\
2               & 1    & $(n-1)R$                                                          & $T+(n-2)P$                                                       & $4R$    & $T+3P$  & C        \\
2               & 0    & $S+(n-2)S$                                                        & $P+(n-2)P$                                                        & $4S$    & $4P$    & D        \\
1               & -    & $(n-1)R$                                                          & $(n-1)P$                                                         & $4R$    & $4P$    & C        \\ \bottomrule
\end{tabular}%
}
\caption{The Table summarised the Expected Utility for each action at each position in the sequence along with the corresponding predicted choice.}
\label{tab:decisions_gm_append}
\end{table}
\newpage
\section{A model of social preferences}\label{sec:appendix_cr}
Following \citet{charness2002} we assume that the utility of the players depends on both their own and others' earnings. In particular, letting $\pi_A$ and $\pi_B$ the  monetary payoffs for players A and B respectively, the utility of Player B  is given by the piecewise linear function:

\begin{align}
     U_B(\pi_A,\pi_B)
    \begin{cases} 
        (1-\rho)\pi_B+\rho \pi_A & \text{if }  \pi_B> \pi_A ~ \text{~(advantageous inequality)}\\
        (1-\sigma)\pi_B+\sigma \pi_A & \text{if }  \pi_B< \pi_A ~ \text{~(disadvantageous inequality)}.
    \end{cases}
\end{align}

with $\sigma<0<\rho<1$ parameters that represent positive and negative reciprocal preferences. This formulation says that the utility of Player B, is a weighted sum of her own monetary and the others player's payoff and may depend on whether A is getting  a higher or lower payoff. Our main objective is to model the behaviour of the \textit{conditional cooperator} type, the one who cooperates when she observes cooperation from at least one past predecessor, and defects otherwise.   We provide the optimal choice for two specifications of this model. First, we consider a modified G\&M model, where the agent follows the same thought process as in G\&M but has also social preferences. That is, this player cooperates when she observes a full cooperation sample and defects if the sample contains only defection. The player expects others to defect when they observe partial cooperation (only one player in the sample cooperated) but we derive the conditions under which this player will cooperate under partial cooperation. In the second specification, we explicitly model the behaviour of the conditional cooperator as one who always cooperates when there is a sample of partial cooperation and defects otherwise (and she also expects others to do so).  In what follows, we consider the expected utility of the players in all the positions and all the potential information conditions on past cooperation. The beliefs of the player's position in the sequence are formed in a identical way as those in \citet{gallice2019co} 

\subsection{Social preferences in a modified G\&M model}
We present the optimal decision for a player at each position and each information set they may receive during the experiment. We denote by $\widetilde{X}$ the untransformed monetary reward. 
\begin{description}

    \item[Position>2, \(m_c=2\)] 
    With a full sample of previous cooperating actions and position uncertainty, the player knows that playing $C$ will motivate the next player in the sequence to also play $C$, while playing D will lead to defection from the next player. The corresponding payoffs are given by:
    \begin{align*}
        U_c &= 4\widetilde{R} = 4R \\
        U_d &= 3\widetilde{T}+\widetilde{P}\\
        &=3((1-\rho)T+\rho S)+P.
    \end{align*}
    The player cooperates whenever $U_c\geq U_d$ which holds for values of $\rho$ that satisfy:
    \[\rho\geq \frac{3T+P-4R}{3(T-S)}\]

    \item[Position>2, \(m_c=1\)] 
     When observing only one prior cooperative action, two potential scenarios arise regarding the sample that the next player will observe. Let us focus on the cooperation decision of a player placed at position $t$ in the sequence. 
     
     \textbf{Case 1:}  If the observed cooperative action originates from the immediately preceding player at position $t-1$, the cooperation of the player at $t$ will generate a complete sample for the next player at position $t+1$. This, in turn, prompts cooperation from the player at $t+1$. The resulting expected utility from this sequence of actions would be $3\widetilde{R}+\widetilde{S}$. 

     \textbf{Case 2:} If the observed cooperative action comes instead from the player at position $t-2$, then the cooperation of the player at $t$ will result in a sample containing only one cooperative action for the player at $t+1$. This would lead to defection by the player at $t+1$. The corresponding expected payoff would be $2\widetilde{R}+2\widetilde{S}$. We assume that the player considers each of the cases equally likely and, therefore, attaches a weight of 1/2 to each. Defection on the other hand, would trigger defection for all the subsequent players. The expected payoffs are given by:

    \begin{align*}
        U_c &= \frac{1}{2}(2\widetilde{R}+2\widetilde{S})+\frac{1}{2}(3\widetilde{R}+\widetilde{S})\\
        &=\frac{5}{2}\widetilde{R}+\frac{3}{2}\widetilde{S}\\
        &=\frac{5}{2}R+\frac{3}{2}((1-\sigma)S+\sigma T)\\
        U_d &= 2\widetilde{T}+2\widetilde{P}\\
        &=2((1-\rho)T+\rho S)+2P.
    \end{align*}
    The player chooses the cooperating action whenever:
    \[\sigma \geq \frac{(2((1-\rho)T+\rho S)+2P-5/2R-3/2S)}{3/2(T-S)}\]
    \item[Position>2, \(m_c=0\)] 
    A player observing a sample of zero cooperating actions can infer that no-one in the sequence before her has cooperated and also she has no way to incentivise the subsequent player to do so. 
    \begin{align*}
    U_c&=4\widetilde{S}=4((1-\sigma)S+\sigma T)\\
    U_d&=4\widetilde{P}
    \end{align*}
    The player cooperates when the condition below holds:
    \[\sigma\geq\frac{P-S}{T-S}\]
    \item[Position=2, \(m_c=1\)] 
    When the player is at position 2 and observes the  cooperating action from the first player, she can then trigger cooperation or defection for all the subsequent players based on her choice. That is:
    \begin{align*}
    U_c&=4\widetilde{R}\\
    U_d&=\widetilde{T}+3\widetilde{P}=(1-\rho)T+\rho S+3P
    \end{align*}
    The player chooses $C$ when:
    \[\rho \geq \frac{T+3P-4R}{T-S}\]

    \item[Position=2, \(m_c=0\)] 
    As in the positional uncertainty case, a player observing a sample of zero cooperating actions can infer that no-one in the sequence before her has cooperated and also she has no way to incentivise the subsequent player to do so. 
    \begin{align*}
    U_c&=4\widetilde{S}=4((1-\sigma)S+\sigma T)\\
    U_d&=4\widetilde{P}
    \end{align*}
    The player cooperates when the condition below holds:
    \[\sigma\geq\frac{P-S}{T-S}\]

     \item[Position=1]
     The first player in the sequence expects that if she cooperates (defects) everyone will cooperate (defect). The expected utility is given by:
         \begin{align*}
    U_c&=4\widetilde{R}=4R\\
    U_d&=4\widetilde{P}=4P
    \end{align*}
    which predicts that the player at the first position always contributes.  
\end{description}

%\begin{table}[H]
%\centering
%\begin{tabular}{@{}cccc@{}}
%\toprule
%Position & Information(\(m_c\)) & $\rho$ & $\sigma$ \\ \midrule
%$>2$ & 2 & $>-0.061$ & - \\
%$>2$ & 1 & - & $>0.333^*$ \\
%$>2$ & 0 & - & $>0.091$ \\
%2 & 1 & $>-2$ &  -\\
%2 & 0 & - & $>0.091$ \\
%1 & - & - & - \\ \bottomrule
%\end{tabular}
%\caption{The Table reports the threshold values for the parameters $\rho$ and $\sigma$ above which a player cooperates, based on the payoffs T=600, R=500, P=100, S=50. \\
%$^*$The value of this parameter is a linear, decreasing function of $\rho$. The reported value is based on the minimum value of -0.061 $\rho$ can take. For $\rho=0.1,~ \sigma=-0.068$, for $\rho=0.2,~ \sigma=-0.318$ and for $\rho=0.3,~ \sigma=-0.568$.}
%\label{tab:parameters_social_gm}
%\end{table}

\subsection{Social preferences in a conditional cooperation model}
This version of the model models the conditional cooperator player, who always cooperates when she observes a sample of partial cooperation (at least one of the previous players cooperated) and expects others to do so as well. The main difference is that there are circumstances where this type cooperates even after observing a full defection sample, aiming to motivate the subsequent players to do so. 
\begin{description}

    \item[Position>2, \(m_c=2\)] 
    With a full sample of previous cooperating actions and position uncertainty, the player knows that playing $C$ will motivate the next player in the sequence to also play $C$, while playing D will lead to defection from the next player. The corresponding payoffs are given by:
    \begin{align*}
        U_c &= 4\widetilde{R} = 4R \\
        U_d &= 4\widetilde{T} \\
        &=4((1-\rho)T+\rho S).
    \end{align*}
    The player cooperates whenever $U_c\geq U_d$ which holds for values of $\rho$ that satisfy:
    \[\rho\geq \frac{T-R}{T-S}\]

    \item[Position>2, \(m_c=1\)] 
     When observing only one prior cooperative action, two potential scenarios arise regarding the sample that the next player will observe. Since by definition a conditional cooperator player cooperates when she observes partial cooperation, the utility of cooperation is given by $\widetilde{S}+3\widetilde{R}$ since only one player defected and the subsequent ones are expected to cooperate. Let us now focus on the defection decision  of a player placed at position $t$ in the sequence that affects the sample that the subsequent players receive. 
     
     \textbf{Case 1:}  If the observed defection action originates from the immediately preceding player at position $t-1$, the defection of the player at $t$ will not dissuade the player at position $t+1$ to cooperate since they will observe a sample of partial cooperation. The resulting expected utility from this sequence of actions would be $3\widetilde{T}+\widetilde{P}$. 

     \textbf{Case 2:} If the observed defection action comes instead from the player at position $t-2$, then defection of the player at $t$ will result in a full defection sample  for the player at $t+1$ which would lead to defection by this player as well. The corresponding expected payoff would be $2\widetilde{T}+2\widetilde{P}$. We assume that the player considers each of the cases equally likely and, therefore, attaches a weight of 1/2 to each.  The expected payoffs are given by:

    \begin{align*}
        U_c &= \widetilde{S}+3\widetilde{R}\\
        &=(1-\sigma)S+\sigma T+3P\\
        U_d &= \frac{1}{2}(2\widetilde{T}+2\widetilde{P})+\frac{1}{2}(3\widetilde{T}+\widetilde{P})\\
        &=\frac{5}{2}\widetilde{T}+\frac{3}{2}\widetilde{P}\\
        &=\frac{5}{2}((1-\rho)T+\rho S)+\frac{3}{2}P.
    \end{align*}
    The player chooses the cooperating action whenever:
    \[\sigma \geq \frac{5/2(T-\rho(T-S))+3P/2-3R-S}{T-S}\]
    \item[Position>2, \(m_c=0\)] 
    A player observing a sample of zero cooperating actions can still cooperate if she expects that she can persuade the subsequent players to cooperate. The corresponding expected payoffs in that case are:
    \begin{align*}
    U_c&=3\widetilde{S}+\widetilde{R}\\
    &=3((1-\sigma)S+\sigma T)+R\\
    U_d&=4\widetilde{P}=4P
    \end{align*}
    The player cooperates when the condition below holds:
    \[\sigma\geq\frac{4P-R-3S}{3(T-S)}\]
    \item[Position=2, \(m_c=1\)] 
    When the player is at position 2 and observes the  cooperating action from the first player, she can then trigger cooperation or defection for all the subsequent players based on her choice. That is:
    \begin{align*}
    U_c&=4\widetilde{R}\\
    U_d&=4\widetilde{T}=4((1-\rho)T+\rho S)
    \end{align*}
    The player chooses $C$ when:
    \[\rho \geq \frac{T-R}{T-S}\]

    \item[Position=2, \(m_c=0\)] 
    At this position the player can affect future cooperation even if she observes defection by the first player. The expected payoffs are given by:
    \begin{align*}
    U_c&=\widetilde{S}+3\widetilde{R}\\
    &=(1-\sigma)S+\sigma T+3R\\
    U_d&=4\widetilde{P}=4P
    \end{align*}
    The player cooperates when the condition below holds:
    \[\sigma\geq\frac{4P-3R-S}{T-S}\]

     \item[Position=1]
     The first player in the sequence expects that if she cooperates (defects) everyone will cooperate (defect). The expected utility is given by:
         \begin{align*}
    U_c&=4\widetilde{R}=4R\\
    U_d&=4\widetilde{P}=4P
    \end{align*}
    which predicts that the player at the first position always contributes.  
\end{description}

%Table \ref{tab:parameters_social_gm} reports the threshold values of the parameters $\rho$ and $\sigma$ above which the player always cooperates. 
%\begin{table}[H]
%\centering
%\begin{tabular}{@{}cccc@{}}
%\toprule
%Position & Information & $\rho$ & $\sigma$ \\ \midrule
%$>2$ & 2 & $>0.182$ & - \\
%$>2$ & 1 & - & $>-0.260^*$ \\
%$>2$ & 0 & - & $>-0.152$ \\
%2 & 1 & $>0.182$ &  -\\
%2 & 0 & - & $>-2.091$ \\
%1 & - & - & - \\ \bottomrule
%\end{tabular}
%\caption{The Table reports the threshold values for the parameters $\rho$ and $\sigma$ above which a player cooperates, based on the payoffs T=600, R=500, P=100, S=50. \\
%$^*$The value of this parameter is a linear, decreasing function of $\rho$. The reported value is based on the minimum value of xxx $\rho$ can take. For $\rho=xx,~ \sigma=-xx$$, for $\rho=xx,~ \sigma=-0.xx$ and for $\rho=xx,~ \sigma=-0.xx.}
%\label{tab:parameters_social_gm}
%\end{table}
%
\subsection{Reciprocal Fairness Equilibrium}
\citet[Appendix~1]{charness2002} propose an alternative model in a multi-player setting where they allow decision makers to express preferences for more general  social-welfare objectives, suggesting the concept of \emph{reciprocal fairness equilibrium}. Let $n$ individuals with material outcomes $\pi_1, \pi_2, \cdots, \pi_n$. The utility function of any individual $i=1,\dots,n$ is a weighted average of their own material payoff $\pi_i$ an a ``disinterested'' social-welfare criterion given by:
\begin{equation}\label{eq:welfare}
W(\pi_1, \pi_2, \dots, \pi_n)=\delta \min[\pi_1, \pi_2, \cdots, \pi_n]+(1-\delta)(\pi_1+\pi_2+ \cdots+ \pi_n)
\end{equation}
with $\delta \in (0,1)$ a parameter measuring the degree of concern for helping the worst-off person  versus maximising the total social surplus. When $\delta=1$ this corresponds to a ``Rawlsian'' criterion, where the objective is to maximise the utility of the least well-off in the society, while with $\delta=1$ the objective is to maximise the total surplus. The utility function of any individual $i=1,\dots,n$ is then given by:
\begin{equation}
    u_i(\pi_1,\pi_2,\dots,\pi_n)=(1-\gamma)\pi_i+\gamma W
\end{equation}
with $\gamma \in (0,1)$. The parameter $\gamma$ measures how much individual $i$ cares about her own interest against the social welfare. When $\gamma=0$ the model collapses to pure self-interest one (in which occasion the model predicts behaviour equivalent to G\&M preferences). Following the same logic as above, Table \ref{tab:welfare}  summarises the expected payoffs for each  player in the sequence as seen from player $i$ in all potential positions and for all potential information conditions (highlighted in the Table). For instance, consider the case when the position is unknown (position>2) and the available information is that none of the previous two players in the sequence cooperated ($m_c=0$). As shown in section \ref{sec:theory}, the decision maker expects  to be at position 4, and given the zero cooperation sample, she may infer that none of the previous players in the sequence has cooperated. Therefore, her expected payoff by playing $C$ is equal to $4S$ while for all the other players, their expected payoff is equal to $T+3P$ consisting of one temptation payoff when they are matched with the player at position 4, and 3 punishment payoffs from the 3 mutual-defect matches. Using the information from each row, it is then straight-forward to calculate the corresponding utility for player $i$ as a convex combination of her own material payoff and the social-welfare criterion. Monte Carlo simulations confirm that our design allows us to identify and successfully recover both $\gamma$ and $\delta$.  
 % Please add the following required packages to your document preamble:
% \usepackage{booktabs}
% \usepackage[table,xcdraw]{xcolor}
% Beamer presentation requires \usepackage{colortbl} instead of \usepackage[table,xcdraw]{xcolor}
\begin{table}[H]
\centering
\begin{tabular}{@{}lcccccc@{}}
\toprule
\multicolumn{1}{c}{\textbf{}}             & \textbf{choice} & \textbf{1}                 & \textbf{2}                   & \textbf{3} & \textbf{4}                    & \textbf{5} \\ \midrule
\textbf{position=1}                       & C               & \cellcolor[HTML]{DDEBF7}4R & 4R                           & 4R         & 4R                            & 4R         \\
\textbf{}                                 & D               & \cellcolor[HTML]{DDEBF7}4P & 4P                           & 4P         & 4P                            & 4P         \\
\textbf{position=2, $m_c=0$}              & C               & T+3P                       & \cellcolor[HTML]{DDEBF7}4S   & T+3P       & T+3P                          & T+3P       \\
\textbf{}                                 & D               & 4P                         & \cellcolor[HTML]{DDEBF7}4P   & 4P         & 4P                            & 4P         \\
\textbf{position=2, $m_c=1$}              & C               & 4R                         & \cellcolor[HTML]{DDEBF7}4R   & 4R         & 4R                            & 4R         \\
\textbf{}                                 & D               & 4S                         & \cellcolor[HTML]{DDEBF7}T+3P & T+3P       & T+3P                          & T+3P       \\
\textbf{position\textgreater{}2, $m_c=0$} & C               & T+3P                       & T+3P                         & T+3P       & \cellcolor[HTML]{DDEBF7}4S    & T+3P       \\
\textbf{}                                 & D               & 4P                         & 4P                           & 4P         & \cellcolor[HTML]{DDEBF7}4P    & 4P         \\
\textbf{position\textgreater{}2, $m_c=1$} & C               & 2R+2S                      & 2R+2S                        & 3T+P       & \cellcolor[HTML]{DDEBF7}2R+2S & 3T+P       \\
\textbf{}                                 & D               & R+3S                       & R+3S                         & 2T+2P      & \cellcolor[HTML]{DDEBF7}2T+2P & 2T+2P      \\
\textbf{position\textgreater{}2, $m_c=2$} & C               & 4R                         & 4R                           & 4R         & \cellcolor[HTML]{DDEBF7}4R    & 4R         \\
\multicolumn{1}{c}{}                      & D               & 2R+2S                      & 2R+2S                        & 2R+2S      & \cellcolor[HTML]{DDEBF7}3T+P  & 3T+P       \\ \bottomrule
\end{tabular}
\caption{The Table   summarises the expected payoffs for each  player in the sequence as seen from player $i$, in all potential positions and for all potential information conditions (highlighted in the Table). Combining the payoffs from each row, the utility of player $i$ can be calculated as the convex combination of her material payoff $\pi_i$ and the social-welfare criterion, defined as a convex combination between the minimum payoff in the sequence and the total surplus.} 
\label{tab:welfare}
\end{table}
\newpage
\section{Simulation}\label{sec:simulation}
In this appendix, we use simulations to explore the robustness of  the estimation method  employed in this paper. Following \citet[Appendix E]{dalbo2019}, we conduct an extensive Monte Carlo simulation to study the ability to identify the prevalence of different strategies in the data and, consequently, the existence of various types in the subject pool. We assess the methodology's robustness, subject to our sample size, by successfully recovering the assumed parameters of our structural model.

We simulate data and assume conditions that resemble those to our lab implementation. We generate simulated data from 50 subjects. Each subject is allocated a type based on predetermined proportions $\pi_k$ with $k\in\{gm,alt,coop, free\}$ for the G\&M, the altruist, the conditional cooperator, and the free-rider, respectively. In every round, subjects are matched in groups of 5 and are allocated to the sequence in a random order. Choices are simulated based on the planned experimental design. That is, each subject provides their conditional choice, dependent on their type and their position in the sequence (strategy method). The process is repeated for 10 rounds, where subjects are matched to a new 5-member group. We report the results of the simulation,  based on the random utility method as presented in the manuscript. 

In this simulation we assume four data generating processes:  the G\&M type, conditional cooperator type modelled assuming  \citet{charness2002} preferences, the free-rider and the altruist. There are 7 parameters to estimate, the mix probabilities for the G\&M, the conditional altruists and free-rising subjects $\pi_{gm}, \pi_{coop}$, and $\pi_{free}$, respectively,  the  positive and negative reciprocal parameters for the conditional cooperator $\rho$ and $\sigma$, the precision  parameter $\beta$ and the tremble parameter $\omega$. We assume again three levels of noise (low, medium and high) and two levels for the reciprocal parameters with $\rho \in [0.100,0.500]$ and $\sigma \in [-0.100,-0.400]$. The percentage of G\&M subjects is set to 40\%, C\&R 30\%, free-riders 20\% and altruists 10\%\footnote{We explored numerous frequency combinations for the four types, and our findings consistently affirm that qualitative results remain unchanged, regardless of the assumed frequency combination.}. We report the results assuming a medium level of noise with $\beta=0.500$ and $\omega=0.15$.\footnote{The level of noise is selected in a way that the simulated data have a success rate between 77 and 80\%.} The results of the Monte Carlo simulation, based on 100 iterations,  are reported in Table \ref{tab:sim2}. For all the parameters, the mean estimate is very close to the assumed value generating unbiased and precise estimates. The simulations reveal that the structural model can recover a wide range of parameters and discriminate between the different types with a remarkably high accuracy. Particularly, in our framework where  discrimination between models depends on slight differences in choice patterns.

\begin{table}[H]
\centering
\begin{tabular}{@{}lccccccc@{}}
\toprule
 & $\pi_{gm}$ & $\pi_{coop}$ & $\pi_{free}$ & $\sigma$ & $\rho$ & $\beta$ & $\omega$ \\ \midrule
True value & 0.400 & 0.300 & 0.200 & -0.100 & 0.500 & 0.500 & 0.150 \\
Estimated value & 0.397 & 0.304 & 0.200 & -0.116 & 0.523 & 0.506 & 0.150 \\
s.d. & 0.025 & 0.022 & 0.009 & 0.063 & 0.119 & 0.070 & 0.017 \\
 &  &  &  &  &  &  &  \\
True value & 0.400 & 0.300 & 0.200 & -0.400 & 0.100 & 0.500 & 0.150 \\
Estimated value & 0.398 & 0.304 & 0.199 & -0.407 & 0.109 & 0.515 & 0.150 \\
s.d. & 0.073 & 0.071 & 0.010 & 0.166 & 0.082 & 0.084 & 0.018 \\
 &  &  &  &  &  &  &  \\
True value & 0.400 & 0.300 & 0.200 & -0.400 & 0.500 & 0.500 & 0.150 \\
Estimated value & 0.396 & 0.306 & 0.199 & -0.486 & 0.566 & 0.510 & 0.148 \\
s.d. & 0.036 & 0.034 & 0.010 & 0.280 & 0.218 & 0.069 & 0.017 \\
 &  &  &  &  &  &  &  \\
True value & 0.400 & 0.300 & 0.200 & -0.100 & 0.100 & 0.500 & 0.150 \\
Estimated value & 0.392 & 0.309 & 0.200 & -0.117 & 0.110 & 0.525 & 0.150 \\
s.d. & 0.091 & 0.090 & 0.009 & 0.085 & 0.059 & 0.107 & 0.018 \\ \bottomrule
\end{tabular}
\caption{Mean estimates from the Monte Carlo simulation for the medium noise ($\beta=0.500, \omega =0.150$). $\pi_{gm}$ is the mixing probability (proportion of subjects) characterised by \citet{gallice2019co} preferences. $\rho$ and $\sigma$ are the distributional preferences parameters of the \citet{charness2002} model. s.d. is the standard deviation. The results are based on 100 iterations.} 
\label{tab:sim2}
\end{table}

Now we repeat this simulation by using the reciprocal fairness model of   \citet{charness2002} as the conditional altruist type, on top of the G\&M, the free-rider and the altruist. Again, there are 7 parameters to estimate, the mix probabilities for the G\&M, the conditional altruists and free-rising subjects $\pi_{gm}, \pi_{coop}$, and $\pi_{free}$, respectively,  the  reciprocal fairness weights $\gamma$ and $\delta$  for the conditional cooperator, the precision  parameter $\beta$ and the tremble parameter $\omega$. We assume again three levels of noise (low, medium and high) and two levels for the reciprocal parameters with $\gamma \in [0.300,0.700]$ and $\delta \in [0.200,0.600]$. The percentage of G\&M subjects is set to 40\%, C\&R 30\%, free-riders 20\% and altruists 10\%\footnote{We explored numerous frequency combinations for the four types, and our findings consistently affirm that qualitative results remain unchanged, regardless of the assumed frequency combination.}. We report the results assuming a medium level of noise with $\beta=0.500$ and $\omega=0.15$.
Table \ref{tab:sim3} summarises the results of the simulation confirming that the experimental design and the estimation method are suitable to identify behavioural strategies and to successfully recover the parameters our statistical model. 
\begin{table}[H]
\centering
\begin{tabular}{@{}lccccccc@{}}
\toprule
 & $\pi_{gm}$ & $\pi_{calt}$ & $\pi_{free}$ & $\gamma$ & $\delta$ & $\beta$ & $\omega$ \\ \hline
True value & 0.400 & 0.300 & 0.200 & 0.300 & 0.600 & 0.500 & 0.150 \\
Estimated value & 0.390 & 0.310 & 0.200 & 0.309 & 0.584 & 0.510 & 0.150 \\
s.d. & 0.055 & 0.059 & 0.010 & 0.099 & 0.176 & 0.077 & 0.019 \\
 &  &  &  &  &  &  &  \\
True value & 0.400 & 0.300 & 0.200 & 0.700 & 0.600 & 0.500 & 0.150 \\
Estimated value & 0.406 & 0.296 & 0.202 & 0.704 & 0.571 & 0.510 & 0.151 \\
s.d. & 0.020 & 0.088 & 0.010 & 0.167 & 0.153 & 0.070 & 0.020 \\
 &  &  &  &  &  &  &  \\
True value & 0.400 & 0.300 & 0.200 & 0.700 & 0.200 & 0.500 & 0.150 \\
Estimated value & 0.406 & 0.316 & 0.202 & 0.733 & 0.183 & 0.505 & 0.151 \\
s.d. & 0.018 & 0.071 & 0.011 & 0.180 & 0.171 & 0.069 & 0.020 \\
 &  &  &  &  &  &  &  \\
True value & 0.400 & 0.300 & 0.200 & 0.300 & 0.200 & 0.500 & 0.150 \\
Estimated value & 0.405 & 0.303 & 0.201 & 0.310 & 0.198 & 0.511 & 0.151 \\
s.d. & 0.024 & 0.064 & 0.010 & 0.064 & 0.137 & 0.071 & 0.018 \\  \bottomrule
\end{tabular}
\caption{Mean estimates from the Monte Carlo simulation for the medium noise ($\beta=0.500, \omega =0.150$). $\pi_{gm}$ is the mixing probability (proportion of subjects) characterised by \citet{gallice2019co} preferences. $\gamma$ and $\delta$ are the reciprocal fairness parameters of the \citet{charness2002} model. s.d. is the standard deviation. The results are based on 100 iterations.} 
\label{tab:sim3}
\end{table}
\end{appendices}

\end{document}